\documentclass{IEEEtran}
\usepackage{cite}
\usepackage{amsmath,amssymb,amsfonts}
\usepackage{algorithmic}
\usepackage{graphicx}
\usepackage{textcomp}
\usepackage{xcolor}
\usepackage{dblfloatfix}
\usepackage{enumitem} 
\def\BibTeX{{\rm B\kern-.05em{\sc i\kern-.025em b}\kern-.08em
    T\kern-.1667em\lower.7ex\hbox{E}\kern-.125emX}}
\begin{document}
\title{
Scaling 
a Blockchain-based Railway Control System 
Prototype 
for Mainline Railways: a Progress Report}

\author{
	\IEEEauthorblockN{Michael Kuperberg\\}
	\IEEEauthorblockA{
		\textit{Blockchain and Distributed Ledgers Group} \\
		\textit{DB Systel GmbH}\\ 
		Frankfurt am Main, Germany \\
		michael.kuperberg@deutschebahn.com
	}
}

\maketitle

\begin{abstract}
Railway operations require control systems to ensure safety and efficiency, and to coordinate infrastructure elements such as switches, signals and train protection. 
To compete with 
the traditional approaches to these systems, a blockchain-based approach has been proposed, 
with the intent to build a more resilient, integrated and cost-efficient system. 
Additionally, the developed blockchain-based architecture enables to run safety-relevant and security-focused business logic on off-the-shelf platforms such as cloud, rather than on specialized (and expensive) secure hardware. 
After implementing a prototype of the blockchain-based railway control system, scaling the approach to real-world mainline and branch operations required a thorough validation of the design choices. 
In this technical report, we show how performance calculations, long-term technology perspectives and law-mandated norms have impacted the architecture, the technology choices, and the make-buy-reuse decisions. 
\end{abstract}

\section{Introduction and Problem Statement}
\label{Introduction}
Railway operations have a complex framework of rules and regulations, with powerful and complex systems in place to control trains, personnel and infrastructure elements. 
These systems have reached a very high pedigree, but their cost and complexity impact competitiveness and operational stability of railways, at a time where zero-emission vehicles, self-driving algorithms, ridesharing/carsharing and changing travel patterns increase the pressure to become more cost-effective. 
The adoption of cutting-edge 
technologies such as ETCS~\cite{etcs-official} and ERTMS~\cite{ertms-official} is progressing slowly, while many secondary railway lines with outdated control infrastructure are eager to upgrade.  

The Blockchain-based Railway Control System (BRCS) \cite{KuperbergKindlerJeschke2020} is an approach which 
\textcolor{black}{
pursuits the vision of automating traffic management and (re-)dispatching, building on software-defined safety mechanisms, train control, interlocking and train protection. 
Innovative technologies such as AI or 5G can be used to support the implementation.} 

\textcolor{black}
{In a \textit{first} realization phase, BRCS
seeks to satisfy the secondary lines'} 
needs by providing a streamlined ``safety core'' for the \textit{minimum} needs of train operations: train collision avoidance, interlocking control, signal control, interfacing automated Train Protection (ATP) such as PZB etc. 
At the same time, the BRCS approach addresses operational efficiency by combining the traffic management aspects (incl. advance scheduling) with ``live'' aspects such as train localization, dispatcher activities, train driver interfaces, and others. 
Through a modularized design and prepared for interfacing EULYNX-enabled components (once those become available on the market), BRCS addresses another pain of railway hardware/software: given the very long expected lifecycles (usually 20-30 years), it is imperative to avoid hardware lock-in or even vendor lock-in. 

After creating a demonstrator and a prototype, scaling BRCS to a ``full'' railway was the next logical step. 
In this paper, we report on our experiences in transitioning BRCS from prototype to a scalable architecture, and we detail how creating the design and the implementation of BRCS is impacted by the lack of modularized hardware/software for infrastructure elements. 
Additionally, we discuss the role of Safety Integrity Layer (SIL) requirements and of the CENELEC\mbox~norms, and why layered software architectures with components sourced from different vendors are challenging in this context. 

\textcolor{black}{During 2020, a significant part of BRCS work efforts were dedicated to creating user interfaces and validating them by surveying potential end users (such as train drivers, dispatchers etc.). 
Also, real-world topological data from GIS systems was imported into BRCS and used for trip planning, monitoring and end-user GUIs. 
For a real-world COTS switch hardware, a controller was set up and corresponding software has been written so that the switch can be controlled over conventional IP networks, from remote locations. 
However, the discussion of these aspects is outside the scope of this technical report.}

This paper is structured as follows: %
Sec.~\ref{Foundations} presents the foundations, 
Sec.~\ref{PotentialCustomers} explains the potential market and customers for BRCS 
and Sec.~\ref{Decentra-Peer-to-Peer} discusses the roles of decentralization and peer-to-peer approaches for BRCS. 
Sec.~\ref{FabricConsensus} explains how consensus algorithms in Hyperledger Fabric work, and how they enable collaborative decision-making following business policies. 
Sec.~\ref{Partitioning} explains why network partitioning is a relevant threat that BRCS must protect against, 
Sec.~\ref{Deletion} discusses the importance of containing data growth and the necessity of being able to delete data from a ledger, and 
Sec.~\ref{Workload} provides approximations of ledger workloads which can be expected in three deployment scenarios. 
Sec.~\ref{ReevaluatingTheChoice} explains why we 
decided to switch to Hyperledger Fabric, 
Sec.~\ref{ProgrammingLanguage} discusses which programming languages were used for smart contracts, other middleware/backend components and for the frontend. 
Sec.~\ref{SafetyIntegrityLayers} presents the impact of the Safety Integrity Layer (SIL) approach on the BRCS architecture, 
Sec.~\ref{Virtualizing} shows how BRCS could benefit from virtualizing the trackside equipment for train localization and train control, 
Sec.~\ref{QuickReleaseCycles} discusses the aspects quick release cycles, 
and 
Sec.~\ref{Conclusions} concludes. 

\section{Foundations}
\label{Foundations}
In the following, we assume that the reader is familiar with enterprise-grade blockchain technology (see e.g. \cite{holbrook2020book,lorne2020book}) and with the general operating terms of mainline railways 
(see e.g. \cite{maschek2018buch} 
and \cite{pachl2018book}). 

\begin{figure*}
	[tbhp]
	\begin{center}
		\includegraphics[trim = 1mm 40mm 46mm 15mm, clip, width=0.98\textwidth]
		{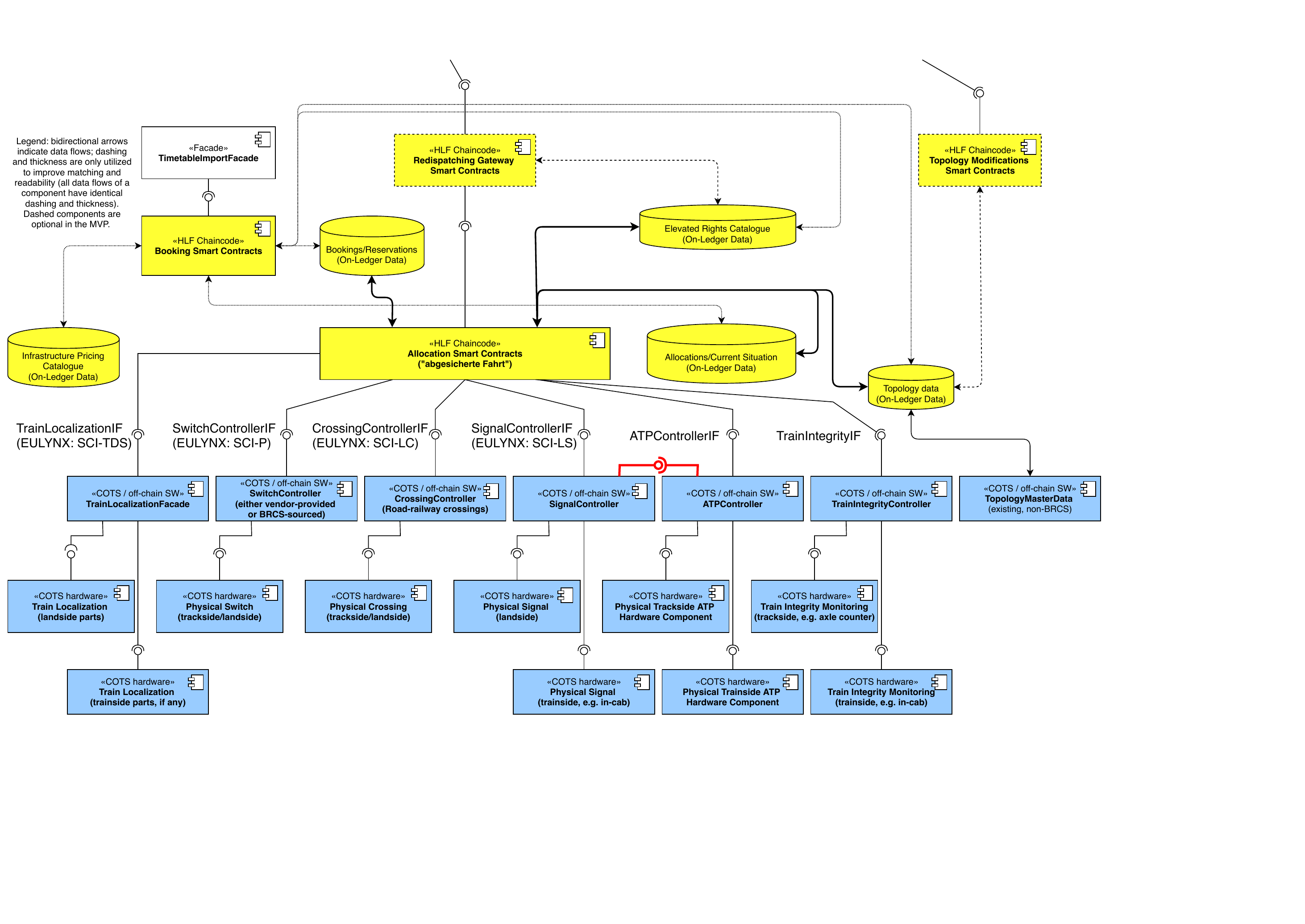}
		\caption{The DLT layer (yellow) and the layer containing 
			pre-ETCS COTS components 
			(blue) of the BRCS architecture. For further layers and components, see the overall architecture in \cite{kuperberg2020ledger}. Interoperability/compatibility with ETCS and the RCA (Reference CCS Architecture) are not shown in the diagram.}
		\label{fig-brcs-architecture}
	\end{center}
\end{figure*}

We also assume that the reader is familiar with 
the architecture of BRCS (shown in Fig.~\ref{fig-brcs-architecture}) and with its background, 
as presented in \cite{KuperbergKindlerJeschke2020}. 

The motivation for BRCS (see Sec.~\ref{PotentialCustomers} for more on stakeholders and customers) originates in the operations challenges: conventional, staffed interlockings provide limited operating times, especially on many secondary networks – and their interlocking hardware is often about to be replaced. 
At the same time, many network operators want simple, IT-supported processes for route bookings and spontaneous journeys, with short processing times and integration into existing systems.

The availability of ETCS does not adequately solve this entire problem, especially since ETCS requires high investments (\textcolor{black}{
replacing control centers or expensive upgrades/retrofits}), which are rarely economical 
on secondary routes. 
ETCS also involves substantial planning and deployment activities, and the installation often requires interruptions or service reductions. 
In addition, there are the costs of trainside ETCS enablement of older rolling stock (see Sec.~\ref{Virtualizing}), if such upgrading is technically possible at all. 
The ideal option would therefore be to maintain as much as possible of the trackside and trainside infrastructure, in conjunction with an efficient software solution for control technology and for the safety core.

BRCS is a combined system that is designed for approval by the EBA in Germany (and similar authorities in other jurisdictions), providing the following core features:
\begin{itemize}
	\item 
a safe interlocking core that is implemented in software and that controls switches, signals, PZB90 elements etc. and thus ensures safety for the operations and the trains
	\item 
management and trusted storage for route bookings (both regular timetables and spontaneous journeys) and the functionality of both a control system and a traffic management
system
	\item 
enablement of train drivers to book routes independently and to start journeys via a mobile tablet application
	\item 
enablement of all actors (RUs, IUs, regulators) to gain a detailed live view of traffic, including through mobile and generic devices
	\item 
a system architecture \textcolor{black}{
that is compatible with installed/legacy technology and field elements but which is also} 
geared to future compatibility with ETCS 
using the open EULYNX interfaces
\end{itemize}

BRCS relies on a distributed and replicated software solution for the interlocking core, which can also run in a cloud environment – so that expensive proprietary hardware is no longer needed. 
In order to efficiently build the BRCS implementation with features such as tamper proofness, resilience, high availability and self-healing, BRCS relies on the proven and high-performance blockchain technology Hyperledger Fabric \cite{androulaki2018hyperledger} (a permissioned DLT for setting up consortial networks), which was created by IBM and has proven itself in high-profile public projects such as TradeLens \cite{tradelens,jensen2019tradelens}. 
Sec.~\ref{ReevaluatingTheChoice} elaborates on the choice of Fabric over 10+ other DLT products, and on the criteria used for this technology decision. 

Usage of Blockchain technology for BRCS means that  
\begin{enumerate}
	\item 
Decisions and underlying logic are stored in a tamper-proof manner; Consensus approvals are stored on the ledger and cryptographically secured. 
	\item 
The blockchain network is managed jointly and transparently and the safety-relevant control logic can only be updated according to consensus (for example, the EBA or any other mandated authority can participate in the network as a supervisor). 
	\item 
Each of the blockchain nodes has the same authoritative "truth": interlocking failures are thus much less likely. 
\end{enumerate}

In addition, BRCS has the following unique selling points:
\begin{enumerate}[label=\alph*)]
	\item 
All information (timetables, route bookings, ongoing journeys) can be seen in one system which unifies all the aspects. 
	\item 
The BRCS implementation can be used directly for billing (routes, station stops), especially for pay-as-you-go scenarios (cf. \cite{kuperberg2018stationshalte}). 
	\item 
The driver can directly see the complete situation on the route network without having to contact the control centre (via phone or radio): the person in the driver's cab is no longer in a ``blind flight''. 
	\item 
BRCS is designed for achieving better safety on routes without ATP or even without signalling, by using software-side warning mechanisms and additional localization technologies  (GPS/GNSS) as an addition to human communication.
	\item 
The BRCS architecture is geared towards independent route booking by vehicles (``headless operations'') and offers interfaces for external or manual traffic optimization and scheduling, e.g. using AI (machine learning) or by human dispatchers. 
\end{enumerate}
\textcolor{black}{
For the Command/Control/Signalling domain (CCS), the EULYNX initiative has created the Reference CCS Architecture (RCA) \cite{rca-eulynx}, targeting infrastructure managers and making use of the data transmission technology defined in ERTMS~\cite{ertms-official}. 
RCA explicitly targets ETCS rather than other contol technologies; only limited backward compatibility with pre-ETCS technology is envisioned (for migration purposes, SCI-LS for light signals is included in RCA).  
As of March 2021, RCA is in the development stage, and therefore it appears no full implementation or even deployment exists yet. 
It remains to be seen how RCA will ensure compatibility with EU-wide \textit{Technical Specifications for Interoperability} (TSI).
} 
%

\textcolor{black}{
In the BRCS project, we already consider the EULYNX SCI interfaces as the target protocol for interfacing field elements - however, many pre-existing field elements do not have controllers or software capable of EULYNX or upgradeable to it. 
Specifically, we intend to support 
SCI-P (for points),  
SCI-LC (for level crossings),
SCI-TDS (train detection systems)
SCI-LS (trackside light signals). 
}

\textcolor{black}{
Regarding RCA, BRCS has focused on prototyping the \textit{Safety Logic} (SL) in the \textit{Safety Control Layer}. 
RCA aspects that are beyond the core value proposition of the BRCS approach, such as 
\textit{ATO Execution}, 
\textit{ATO Transactor}, 
\textit{ATO Vehicle}, 
\textit{Mobile Object Locator} (referring to construction equipment etc.), 
\textit{Person Supervisor and Locator}, and others
are on our long-term agenda (as we monitor the RCA development progress). 
}

\textcolor{black}{
On the other hand, BRCS architecture includes physical trackside (landside) signaling whereas RCA does not: RCA only applies to CCS technology without such signaling, i.e. only with in-cab signalling (if any). 
Therefore, BRCS cannot implement RCA 1:1. 
Consequently, as one of our next steps, we plan to align the BRCS architecture \textit{terms} (cf. Fig.~\ref{fig-brcs-architecture}) with those of RCA. 
For example, RCA distinguishes between PAS (\textit{planning system}) and PE (\textit{plan execution}, in the \textit{Plan Implementation Layer}). 
}
 
\textcolor{black}{
In the context of EULYNX and RCA, OCORA \cite{ocora} is the Open CCS On-board Reference Architecture. 
As of March 2021, it is in the so-called Gamma status (post-beta, pre-release). 
BRCS is monitoring the activities within OCORA, but with our current focus on the safety core, OCORA currently has no impact on our activities. 
}

\textcolor{black}{
Technology-wise, the German national digitalization program Digitale Schiene Deutschland (DSD) \cite{dsd} starts with ETCS and ATO. 
Regarding ETCS, DSD prioritizes heavily-used lines (such as TEN corridors) and networks (such as the newly built Stuttgart21 infrastructure). 
It also performs research on 5G, AI-based traffic management and dispatching, 
and on combined train positioning (using Lidars, image recognition and other techniques). 
Overall, DSD and BRCS have different focus groups and they complement each other: while DSD prioritizes further 
strategic improvements on main lines, the BRCS approach prioritizes 
tactical short-term 
improvements on branch lines and on low-traffic segments. 
Thus, BRCS strives to provide ``quick benefits'' on the lines where ETCS is not planned to be deployed in foreseeable future. 
In addition to aligning our efforts with the DSD objectives, we are also closely monitoring successor activities to Shift2Rail \cite{shift2rail}, and other research and strategic initiatives. 
}

\section{
	Potential Customers and Usage}
\label{PotentialCustomers}
The target market of BRCS is not restricted to Germany or to the European Union, but as the BRCS team is 
part of Deutsche Bahn AG and as Germany is known for its rigorous (and often lengthy) acceptance testing, we decided that Germany is the most appropriate juridical and technological context for our work. 
In particular, we hope that gaining the deployment permission in Germany is very likely to open the doors to BRCS in other countries and legislations. 

To better gear BRCS towards prospective customers, we had to understand those customers' needs and pain points, and to work out the value proposition for BRCS. 
In Germany, railways are deregulated - a large number of train-operating companies (both freight and passenger) operate on a network that comprises subnetworks owned and maintained by infrastructure undertakings. 
Still, the situation is very different from (for example) the United States of America, where \textit{private} freight railway companies own large and small networks, and where state-owned trackage is rare (and mostly found in areas with significant regional and commuter passenger traffic). 
Likewise, the situation is very different from (for example) France where the state railways (SNCF) operate over 99\% of regional train passenger traffic (and mandatory tendering of regional lines will not start until 2024): in Germany, the private competition accounts for over 35\% \cite{db-wettbewerbsbericht-fuer-2018-2019} of regional train passenger traffic as of January 2021. 

In Germany, traditionally, certain special-purpose networks (such as railway lines within the boundaries of maritime ports or large industrial estates) were already owned by those estates. 
However, even lines with intensive passenger traffic are not exclusive to DB Netz (which is the largest infrastructure undertaking): for example, AVG (a state-owned 
operator) maintains a 
network of electrified lines around Karlsruhe. 
Still, all these railways have the same heritage when it comes to signaling and train protection: they use PZB and LZB; 
ETCS rollout is in very early stages with 
320 km 
having ETCS Level~2 and 0 km having ETCS Level~3. 
No ERTMS deployment has been completed in Germany as of January 2021. 

Of the pain points that we have identified from interviews and through own research, the top five are
\begin{enumerate}
	\item Outdated technology in control centers (especially when these are nearing the end of useful life) incurs high operating costs even where the traffic is light, because constant on-premise presence of an employee is required to operate the legacy technology - and recruiting the next generation for these jobs often proves problematic.
	\item Replacing the outdated technology through ETCS is often 
	expensive both for the trains and for the infrastructure, and the costs can be 
	disproportionately high when the impacted lines see little traffic, or when the impacted trains cannot accomodate the equipment without costly adaptations to the train itself.
	\item Introducing ETCS on the infrastructure side while being unable to upgrade \textit{all} potential vehicles on that line with ETCS would mean a costly co-existence of ETCS with previous technology (or, alternatively, blocking out ETCS-less traffic when only ETCS is supported) - interestingly, all trains that operate on the few German lines with operational ETCS have at least two further train protection equipment on board (PZB and LZB, in addition to ETCS).
	\item For historical and compatibility reasons, development of timetabling, billing, dispatching, automated train control, monitoring, signaling, train protection etc. has created an array of disparate IT systems which show a varying degree of compatibility and data exchange. 
	As we have already discussed in \cite{kuperberg2020ledger} and \cite{kuperberg2018stationshalte}, an integrated approach would improve many operational aspects, and simplify on-demand train service and schedule modifications. 
	\item Vendors of control centers, trackside infrastructure, trainside IT equipment etc. are eager to maintain their competitive edge and customer ``loyalty'' and less eager to maximize interoperability, to agree on open interfaces or to sell individual modules (rather than integrated offerings). 
	For example, despite the multi-national EULYNX\mbox~effort which started in 2014, we have not found a single offering of, for example, switch that would support the EULYNX-defined interfaces. 
	Likewise, it is acknowledged \cite{liefferinge2017-etcs-compatibility} that ETCS does not automatically mean compatibility. 
	Unfortunately, the openETCS \cite{karg2016openetcs} approach has not triggered a significant level of vendor interoperability. 
\end{enumerate} 

Building on these pain points, the best potential for BRCS' \textcolor{black}{
first implementation stage} 
can be seen in \textit{reusing} suitable pre-existing trainside and trackside equipment (e.g. PZB) while creating new software that satisfies as many demands as possible with the focus on the train operation. 
BRCS should thus focus on the two core steps: booking a train ride and executing a train ride (incl. train protection, monitoring and signaling), while consuming data (e.g. precomputed timetables) and exposing the data (e.g. completed rides) that is dealt with by other systems. 
BRCS would be particularly attractive to branch lines seeing moderate traffic, though we believe that the concept and appear of BRCS can appeal to larger networks, too - however, with ETCS engraved in laws 
and in long-term financial planning, we have no illusions that BRCS' \textcolor{black}{
first stage} could 
\textcolor{black}{
supplant} 
ETCS. 

Beyond conceptual advantages, the financial viability of BRCS is impacted by the ``economies of scale'' that finalized and deployed solutions such as ETCS expose: we assume that the development budget for ETCS has already been spent, and is refinanced through finalized wide-scale sales; additionally, EU funds for ETCS equipment have been set up and are being dispersed. 
BRCS as a ``newcomer'' has to acquire seed budget to develop a cost-effective solution even though its team has no pre-existing reusable assets 
from developing or marketing such a system. 
With ETCS efforts going back to 1991, BRCS must be focused and agile to create a solution that finds buyers and supporters. 

Companies producing competing control technology include industry heavyweights such as Alstom/Bombardier, Huawei, Siemens and Thales (to name just a few from the EU and from China), and smaller players such as BBR, RDCS and Scheidt\&Bachmann. 
A number of further companies, such as GTB and Pilz, produce trainside and/or trackside equipment, especially for the train protection such as PZB. 

\section{
Decentralization and Peer-to-Peer Aspects of the Architecture}
\label{Decentra-Peer-to-Peer}

Blockchain protocols and products build on peer-to-peer (P2P) networks and algorithms, accounting for delays, connectivity outages and unreliable nodes. 
P2P protocols support technical decentralization, and use communication patterns which do not require Quality-of-Service (QoS) contracts or Service Level Agreements between blockchain network participants. 
Even permissioned DLT implementations use probabilistic protocols: for example, Hyperledger Fabric uses the so-called gossip protocol\cite{berendea2020fair}. 

This \textit{technical decentralization of the setup} comes along with \textit{organizational decentralization of the decision-making} (``consensus''). 
The two aspects of decentralization contrast with the \textit{logical centralization of data} which is a cornerstone principle of the DLT: each participant is enable to obtain and to maintain a full authoritative copy of the ledger - and the data on the shared ledger is the same for all blockchain network participants. 
Under certain circumstances, a participant may experience delays in seeing this uniform, consistent truth,
and certain DLT products may by susceptable to network partitioning (which may lead to multiple truths, see Sec.~\ref{Partitioning}), but in general, all DLT implementations strive to maintain a single consistent (and authoritative) truth even in the presence of the CAP theorem \cite{6133253}. 
Many use cases that utilize DLT advertise decentralization - as the ability to operate without a central authority and in a peer-to-peer fashion. 
When discussing and explaining BRCS, we often faced the necessity to position our architecture in terms of P2P capabilities, and to explain whether decentralization and the P2P principle can actually work for the railway domain: specifically, whether decisions can be negotiated in a P2P way. 

\begin{figure}
	[h]
	\begin{center}
		\includegraphics[trim = 1mm 76mm 150mm 50mm, clip, width=0.7\textwidth]
		{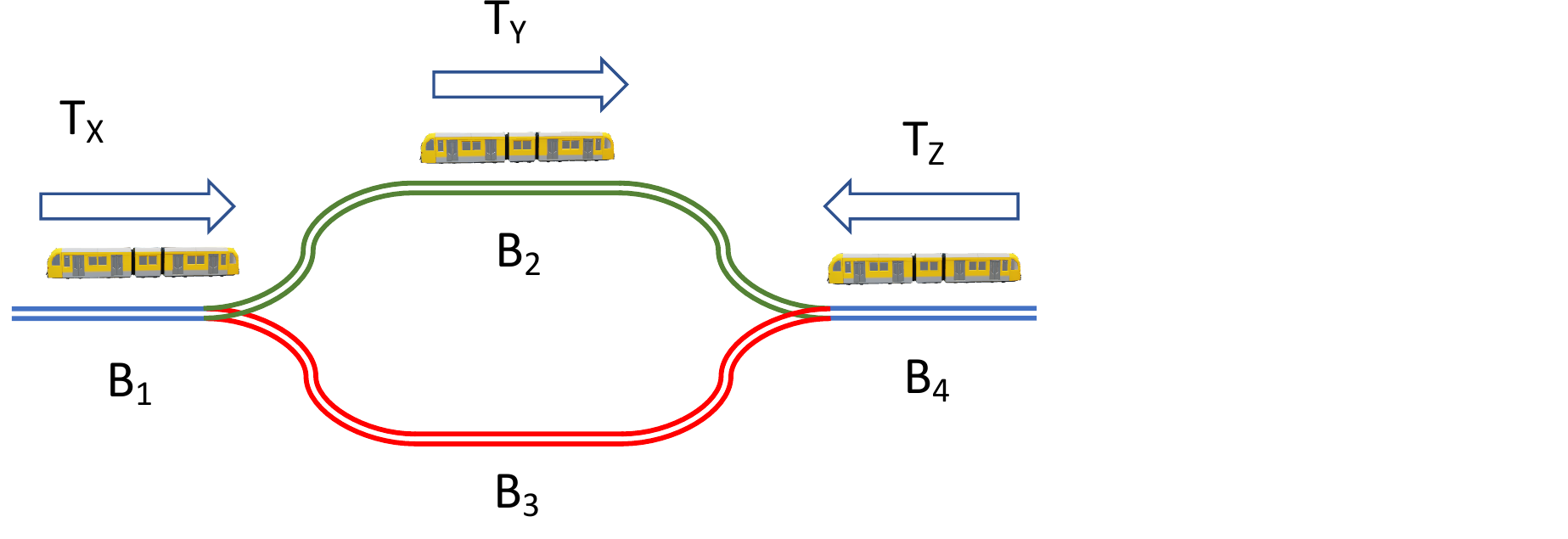}
		\caption{Example situation for exploring the possibibility of a P2P one-to-one coordination ($B$ are blocks and $T$s are trains; signals and switches not shown)}
		\label{gleisplan}
	\end{center}
\end{figure}
To understand the issue of such \textit{business-logic-level P2P} for BRCS, consider a simple single-track 
line 
section
shown in Fig.~\ref{gleisplan} with 
a single 
passing siding, 
and 
without diverging or crossing lines. 
The sections (blocks) are $B_1$ through $B_4$; we do not index the switches. 
To start with, assume that three trains $T_X$, $T_Y$ and $T_Z$ are in 
operation on this line; the arrows show the direction of travel. 
If the trains were to coordinate their movements in a P2P, one-to-one style, 
coordination of $n$ trains would require $(n-1)+(n-2)+...+1=\frac{n\times(n-1)}{2}$ bilateral conversations. 
Note that the one-to-one style does not dictate whether \textit{all} the peers share \textit{one} global common authoritative state or not. 

Such a coordination is a set of complex algorithms which ensure safety, and is particularly challenging because 
the outcome of a naïve bilateral coordination may be provisional until further coordinations are completed: 
for example, in Fig.~\ref{gleisplan}, $T_{X}$ and $T_{Y}$ may come to the agreement that $T_{X}$ may proceed to the lower track ($B_3$) of passing loop 
- however, as such coordination disregards $T_{Z}$'s intentions, this will create a livelock: none of the three trains will be able to move on in the intended direction. 
The livelock will only be detected if the coordination between $T_X$ and $T_Y$ co-analyzes the impact on the other trains in the network - potentially, on \textit{all} other trains in the network, which may be impacted not just immediately after a decision but only after trains have proceeded to a different section of their route. 

Additionally, the coordination between $T_X$ and $T_Y$ may not be the only one executed at a given time, and parallel coordination could be happening. 
The exponential complexity of $\frac{n\times(n-1)}{2}$ may be acceptable for very small networks (such as one in Fig.~\ref{gleisplan}) with low traffic intensity, but its inefficiency is aggravated by the need to manage parallelism. 
For example, $T_Y$ may be concurrently coordinating with $T_Z$ and that coordination may conclude faster, resulting in the agreement that $T_Z$ 
would proceed to the lower segment $B_3$ of the passing siding while $T_Y$ waits. 
Obviously, this outcome does \textit{not} create a livelock (as $T_Y$ can proceed to $B_4$) - however, the agreement between $T_Y$ and $T_Z$ makes the pending in-the-works agreement between $T_X$ and $T_Y$ void and dangerous, as $T_X$ would attempt to move into a section to be occupied by $T_Z$ (when the coordination between $T_X$ and $T_Y$ started, that section was empty!). 
This small example shows that concurrent P2P-style one-to-one coordinations are potentially inefficient (as up to $\frac{n\times(n-1)}{2}$ safety-checking coordinations are necessary) and complex (i.e. necessitating ACID-style transaction managers) or even dangerous when the intersection between the involved sets of resources is non-empty. 

An unsophisticated approach to mediate this situation would be to enforce the consideration of all relevant resources (incl. the liveliness of all other trains) and to serialize all coordinations (removing concurrency and eliminating race conditions). 
This means that any of the coordination-inducing peers must be able to execute the same safeguarding/control logic ensuring the considration of all relevant and concerned actors and resources within the given railway network - however, this raises the questions about capability, integrity, precision, reliability etc. of that P2P peer. 
At this point, what started as business-logic-level-P2P coordination between two peers evolves into a coordination that concerns all peers. 
Effectively, this matches the BRCS architecture where the control logic is a trusted smart contract that can be run by any peer and which is secured through consensus. 
In particular, when implementing BRCS using Hyperledger Fabric, consensus follows a configurable endorsement policy - and endorsement involves re-running the smart contract (by the endorsers, and against the local replica of the ledger) to verify the \textit{transaction proposal} issued by the requester. 

In terms of optimization, ensuring that naïve P2P-style one-to-one coordination would reach a global maximum (rather than a local maximum) would be a daunting task, involving complex rules and multi-step protocols. 
A lot of research exists in this direction, including concepts for Multi-Agent Systems. 
Yet if decentralization would be implemented without mandatory rules that codify the maximization of the overall benefit, each participant would likely seek the maximum of the individual benefit, and the participant which snatches the right to coordinate has more chances in overtaking the others.

Thus, conceptually, BRCS combines both centralized and decentralized aspects in the following manner:
\begin{itemize}
	\item \textbf{Central network-wide truth in data}: the ledger holds the central, shared, unique truth: the system state (e.g. precomputed timetables, ahead-of-time reservations of infrastructure elements and during-trip occupation and state of the infrastructure). 
	\item \textbf{Central network-wide rules in algorithms}: the smart contracts (stored on the ledger) codify the safeguards for train control and dispatching, and are uniform for all business partners; they are defined and deployed collaboratively.  
	\item \textbf{Decentralized governance}: there is no ``master admin'' who can accept or bar participating organizations - business network administration is based on consensus and on transparency.
	\item \textbf{Decentralized transaction approval}: transactions are verified (endorsed) and ordered (serialized) by peers/orderers distributed over the organizations participating in the business network. 
	\item \textbf{Decentralized propagation of transaction proposals and confirmed transactions}: using the gossip protocol, there is no bottleneck or central broker that is responsible for information distribution. 
	\item \textbf{Decentralized infrastructure}: each participating organization sets up its own node software - with proper distribution (e.g. over different cloud providers and connectivity networks), there is no single point of failure. 
\end{itemize} 

\section{Consensus in Hyperledger Fabric}
\label{FabricConsensus}
To explain how decentralized decision-making in Hyperledger Fabric maps to railway control needs, it is important to remember that Fabric is a permissioned ledger which builds business networks from organizations. 
An organization can be a supervisory authority such as EBA, a railway undertaking or an infrastructure undertaking. 
Within a business network, one or several channels exist: a channel is a business context. 
For example, a channel covering infrastructure owned by AVG could be separate from the channel covering the infrastructure owned by SWEG, etc.

Organizations that participate in a channel agree on endorsement policies (the following is a simplified view; refer to the Fabric documentation for an authoritative in-depth coverage):
\begin{itemize}
	\item \textbf{Channel membership endorsement policy}: in the case of BRCS, this policy would insure that consensus governs any changes to the channel's body of organizations. 
	\item \textbf{Chaincode update endorsement policy}: in the case of BRCS, this policy would stipulate that the supervisory authority must explicitly agree to any changes. Which other organizations can (or must) endorse is subject to business agreements and laws/ordinances. 
	\item \textbf{Data update endorsement policy}: in the case of BRCS, this policy would govern any write requests made to the ledger-stored data, except chaincode. Here, explicit endorsement from the government authority would not be needed. 
\end{itemize}

The setup of a channel also includes any rights and privileges to read the ledger-stored data - the core difference is that no consensus is needed for read requests.
Still, the flexibility of Fabric enables the BRCS implementation to restrict data visibility, even down to individual principals within any organization participating in a channel. 
Note that established X.509 PKI cryptography is used to administer these privileges, which opens the possibility of integrating these tasks into pre-existing enterprise CAs. 

Finally, it is important to understand that in Fabric, an organization's voting and endorsement rights do not depend on the number of nodes that the organization sets up. 
Why at least one peer node is necessary to participate (to host a ledger copy) and at least one peer node must run in endorsment mode to permit active consensus participating, 
all organizations are free to choose the number of peer node, depending on their reliability and redundancy needs. 
Fabric is designed to sustain high performance even when the number of peer nodes is high; this is achieved through the gossip protocol and through symmetric setup of peer nodes within an organization. 

\section{
Network Partitioning}
\label{Partitioning}
Consensus must work reliably even when the distributed ledger is experiencing disturbances such as node failures or network outages. 
More specificially, the consensus implementation must prevent a situation where several conflicting data versions are created concurrently. 
At the same time, the network shall continue to operate even when one (or several) nodes fail or cannot be reached. 
As Hyperledger Fabric builds on top of the networking infrastructure, we have to investigate how network outages impact the operations of BRCS' underlying layers. 
Specifically, network partitioning is a form of network outages that is of particular interest since it is a known threat vector \cite{alquraan2018-partitioning} in distributed systems. 

In \cite{kuperberg2020ledger}, we announced that robustness against network partitioning would be one of the guiding factors when seeking alternatives to Ethereum with Proof-of-Work. 
These planned activities were motivated by 
the observation that forks in Ethereum are not just a nasty side effect that results in delays (before a block is trusted), 
but can lead to inconsistencies if a blockchain network splits in two parts that are able to operate independently. 

In a separate paper \cite{kuperberg2020partitioning}, we have analyzed different consensus algorithms in different blockchain implementations for these algorithms' abilities to detect and to prevent network partitioning. 
The algorithms included Proof-of-Work in Ethereum, Aura Proof-of-Authority in Parity Ethereum, and Proof-of-Authority in Hyperledger Fabric; we have also discussed generic Proof-of-Stake and generic Proof-of-Space. 
Our research has shown that of these, only Proof-of-Authority in Hyperledger Fabric is capable of detecting and preventing network partitioning, by configuring the \textit{endorsement policy} following the formula that we have derived and explained in the paper: $2 \times m > a$, where $a$ is the number of authorities and the Fabric endorsement policy is configured as ``at least $m$ out of $a$'' with $m < a$. 

The ability to prevent network partitioning is one of the requirements that we have set up in preparation of the blockchain product re-evaluation, which we will describe in Sec.~\ref{ReevaluatingTheChoice}. 
After  \cite{kuperberg2020partitioning}, we expanded our research on network partitioning to cover 
a 
further blockchain 
product 
and 
its 
consensus algorithms: Quorum (``Enterprise-grade Ethereum'' with permissioning).
Quorum can be configured with different consensus algorithms: Clique PoA, Instanbul BFT 1.0 and 2.0, as well as RAFT / PBFT. 

Clique PoA \cite{clique} is often associated with Ethereum, and is officially described as follows: ``in Clique networks, approved accounts, known as signers, validate transactions and blocks. Signers take turns to create the next block. Existing signers propose and vote to add or remove signers.'' 
The latter is also explained in more detail: 
``A majority of existing signers must agree to add or remove a signer. That is, more than 50\% of signers must execute [the API method] clique\_propose to add or remove a signer. For example, if you have four signers, the vote must be made on three signers.''
Only one signer must sign, so despite Clique's differences to naïve round-robin PoA, Clique is still vulnerable to network partitioning as it cannot detect it (on the consensus level) and also cannot be configured against it. 
The restrictions on \texttt{clique\_propose} mean that in the case of strictly disjunct partitions (two or more), all partitions are not able to add or remove signers (even though signing itself continues to be available). 
However, if a given signer is present in 2 or more partitions in a Clique-running network, that signer \textit{can} enable adding/removing signers in those partitions. 

Instanbul BFT (IBFT 1.0) \cite{ibft10} is specified as an Ethereum Improvement Proposal (EIP) and refers to Clique and to PBFT as inspiration - but it is also used within Quorum (a permissioned, enterprise-grade Ethereum implementation).
The PoA rule within IBFT 1.0 is that in a network with $N=3F+1$ authorities (``validators'', i.e. endorsers), a transaction proposal is considered acceptable if it assembles $2F+1$ or more endorsements, i.e. the IBFT 1.0 consensus can tolerate up to $F$ faulty (``Byzantine'') nodes. 
The validators exchange their endorsements via a gossip protocol to supplement P2P messages; there is no specific orderer role (unlike in Hyperledger Fabric): each validator decides individually whether to accept a transaction proposal, based on its knowledge of $N$ and based on the number of endorsements received over P2P/gossip. 
The choice of ``at least $2F+1$ out of $3F+1$'' means that in the case of a network partitioning where each validator is communicating within one partition only, conflicting truths cannot emerge since at most one partition can include $2F+1$ validators; if none of the partitions includes $2F+1$ or more, all partitions are stalled. 
In the case of Byzantine validators which interact with more than one partition, the ability to achieve consensus in 2 partitions would mean that $F+1$ faulty/Byzantine validators would have to be present (out of $3F+1$): $(3F+1)-(2F+1)=F$ validators remain for the second partition, but that partition requires further $(2F+1)-F=(F+1)$ validators and those must be double-acting, i.e. faulty. 
However, \cite{ibft10} explicitly declares that ``The system can tolerate at most of F faulty nodes in a N validator nodes network, where N = 3F + 1''. 
Thus, it appears that concurrent truths with Byzantine validators in IBFT 1.0 are only possible in a situation which IBFT 1.0 cannot cope with by design. 
Unfortunately, one has to consider the minimum number of validators that makes sense with $N=3F+1$ without rounding (obviously, $N$ cannot be \textit{any} integer number), 
and this is $4$ if we consider $N,F\in \mathbb{N}$. 
To see why this is essential, consider that 
even $N=6$ (which means that $F$ is between 1 and 2) carries the risk of two conflicting truths:
\begin{itemize}
\item assuming $F=1$, for two partitions, each partition can have $2F+1=3$ validators, even without a single Byzantine node
\item assuming $F=1$, for three partitions, two partitions can have $2F+1=3$ validators, involving a single Byzantine node that is present in both (or all three) partitions
\item assuming $F=2$, only one partition can have $2F+1=5$ validators, even when two Byzantine nodes are present in all partitions
\end{itemize}
IBFT 2.0 \cite{2019arXiv190910194S} (also used in the Hyperledger Besu implementation and in Quorum) is an extension of IBFT 1.0 and claims to ``address the safety and liveness limitations'' of IBFT 1.0 - \cite{2019arXiv190910194S} explicitly addresses partitioning, too. 
Both IBFT versions have an important difference from PoA used in Hyperledger Fabric: IBFT does not have the flexibility of Fabric when it comes to defining endorsement policies. 
Specifically, Fabric's PoA allows to enumerate \textit{specific} endorsers (authorities/organizations) as must-endorse, wheras IBFT permits \textit{any} $2F+1$ out of $3F+1$ for a validation to succeed. 

\textcolor{black}{We plan to address RAFT and PBFT in forthcoming publications.} 

\section{Data Growth and Erasability in Consensus-Driven Append-Only DLT and Blockchains Which Have No Built-in Deletion Capabilities}
\label{Deletion}
Another important aspect for BRCS is the ability to operate under realistic workloads. 
However, the ``append-only'' design pattern is prevalent in DLT, and is also found in Ethereum (both \texttt{geth} and Parity as well as in Quorum) and in Hyperledger Fabric. 
This design choice means that data accumulates over time as individual entries (transactions and blocks) cannot be deleted. 
The only realistic choice is to re-initialize a new blockchain structure after some time has passed, or after a predefined amount of data has been accumulated. 
Even such a primitive workaround requires extensive coordination between network operators; it is also quite challenging to ensure that relevant data from the ``past'' blockchain is carried over from the ``old'' blockchain, as \textit{selective} deletion is not possible.

\begin{figure}
	[tbph]
	\begin{center}
		\includegraphics[trim = 7mm 85mm 48mm 48mm, clip, width=0.49\textwidth]
		{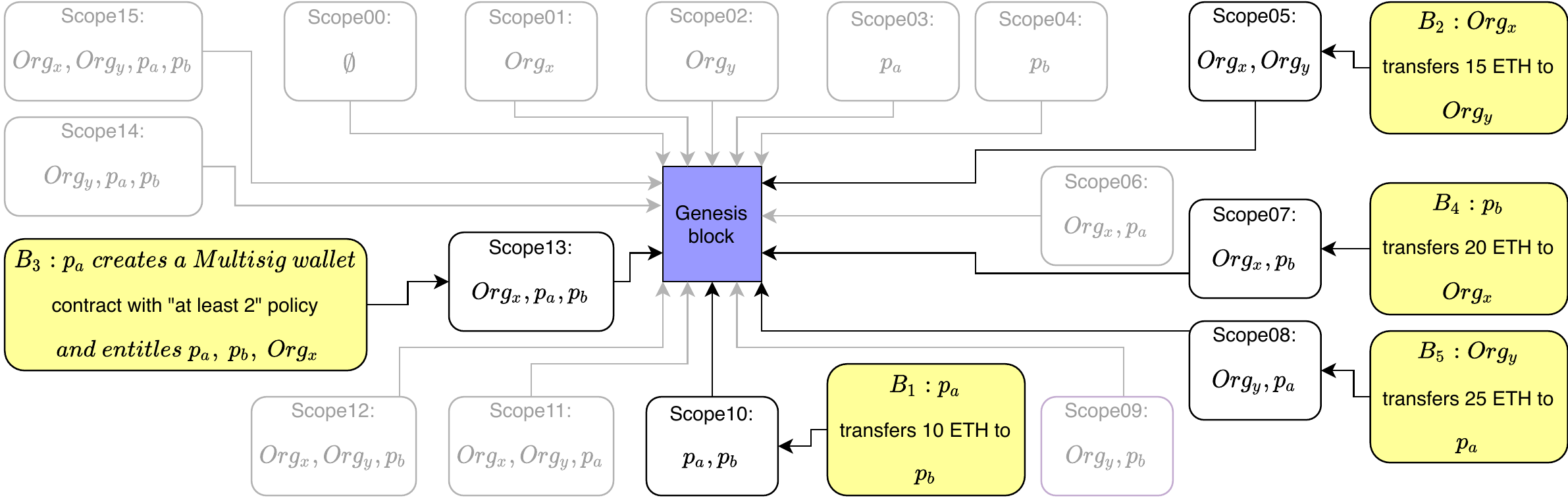}
		\caption{Illustration taken from \cite{kuperberg2020deletion}: non-linear blockchain, transactions with the same context scope are located in the same branch (subtree); subroots shown in grey are not needed right from the outset and can be created later on-demand once blocks with corresponding transactions are mined} 
		\label{fig-context-tree}
	\end{center}
\end{figure}

Many projects help themselves by minimizing on-chain data, while storing the majority of data off-chain: for example, only hashes (fingerprints) of data are stored on-chain. 
Effectively this separation relinquishes some of the promises of blockchain, such as non-repudiation and auditability (since off-chain data can be deleted without impacting on-chain hashes). 
In a separate research paper \cite{kuperberg2020deletion}, we have outlined how a possible solution would look like: we propose using a non-linear data structure (a \textit{context tree}, see example in Fig.~\ref{fig-context-tree}), and a transaction is placed in exactly one of the tree branches based on that transaction content. 
The paper explains how this placement is achieved, why such a structure brings more benefits than downsides, and how it could improve the GDPR compliance of blockchain-based applications. 

In a separate technical report \cite{kuperberg2020arxiv}, we expand the erasability research to DAGs (Directed Acyclic Graphs), which see use in DLT solutions such as IOTA. 
Given the priorities within the BRCS project, we decided not to implement the erasability on our own, but rather to seek community feedback in the first step. 
At the current stage, BRCS can proceed using a DLT that is not capable of deleting; the issues of data growth and of erasability can be addressed at a later stage as it does not require a refactoring of the BRCS architecture. 
Thus, to ensure that our results are not being productized (or plagiarized) by competing entities, we have filed \cite{DE1020182202249} the idea with DPMA (Deutsches Patent- und Markenamt), i.e. with the patent and trademark office in Germany. 
Then, after 
presenting our research at the scientific conference, 
we presented it to the Hyperledger Architecture Working Group. 
As part of our future work, we hope to see the contribution in \cite{kuperberg2020deletion,kuperberg2020arxiv,DE1020182202249} being implemented in a blockchain product. 

\section{Workload Modeling}
\label{Workload}
BRCS must be implemented so that it is able to cope with real-life train frequencies and infrastructure usage. 
Since performance must be considered as early as possible in the development phase, we decided to create 
a performance model of the workloads that BRCS can expect, and combine them with knowledge about DLT performance that has been published. 
Such analysis can help with DLT product choice, design choices, implementation decisions, and with deployment strategies. 
It is also an important input for stakeholder management and for test design. 

To start with, possible workloads on a BRCS implementation were derived from the following three scenarios: 
\begin{enumerate}
	\item a very short (12,8 km) single-tracked branch line which is normally operated by a single shuttle train, resulting in an hourly passenger service in each direction between ca. 5:00 and ca. 21:00; the line has a single switch where it branches off (all passing loops are out of service as of 2020) and 14 level crossings with local roads; freight service is almost nonexistant and the average train speed can be estimated to be 60 km/h
	
	\item a mid-sized regional network (10 lines totalling 397,9 km line length and ca. 750 km track length) with top speeds of 140 km/h, both mainlines and branch lines, partially multi-tracked and partially electrified, served by several train companies (with routes partially starting/ending outside this regional network), and seeing heavy freight traffic
	
	\item a freight corridor section 
	(itself part of a larger EU freight corridor), comprising several mainlines and having a route length of ca. 630 km and a track length of ca. 1600 km with top speeds of 250 km/h, completely electrified, seeing very intensive passenger and freight traffic, and served by several train companies
\end{enumerate}
In scenario 2, 
during the peak hours
the mid-sized network sees more than 30 trains within the network in a window of 60 minutes, up to 34 trains concurrently present over 56 minutes. 
To account for possible delay-induced overloads, we have considered the double number, i.e. that up to 70 trains can be present in the network at the same time. 
In literature, the estimated average block length (between two signals) is often 1000 m, which means that the network is comprised of 750 blocks. 

Scenario 3 has the heaviest train presence, with up to 300 scheduled trains present at the same time in the network. 
For scenario 3, we decided to assume a shorter block length of 700 m, because infrastructure and signaling improvements are carried out to further increase the capacity of that corridor. 

To derive a performance model (based on queuing theory) for each of the three above scenarios, we needed to estimate the arrival rate $\lambda$ and the processing/departure rate $\mu$ (assuming, for simplification, that BRCS will operate in a First-In-First-Out fashion).
In the blockchain layer of BRCS, the write transactions have a comparatively higher duration, whereas the read request can be executed very quickly, especially if there is an index-like side structure as the built-in ``world state'' in Hyperledger Fabric. 
Therefore, we focus the workload quantification on the write transactions, even though the read transactions would have to be considered once a fully-featured performance model will be constructed. 

For the write requests in the above scenarios, we assume that during the train operation and for each traversed block, BRCS issues two writing transactions (on entering the route block and on leaving the block). 
Before the train operation, the route blocks are pre-allocated, which leads to one writing request for each allocation; note that re-allocations are not uncommon due to reroutings, construction, forecastable overloading, etc. 
Therefore, in a setting where both the pre-booking of a train trip and the actual (later) trip are executed within BRCS, we assume 4 writing operations per train per block. 
Write requests that belong to infrastructure elements such as switch operations, signalling, etc. have to be considered separately. 

The formula that we will use is 
trains-per-hour 
$\times$ 
blocks-per-second-per-train 
$\times$
number-of-write-transactions-per-block-per-train. 
Let's assume that by considering, \textit{per block} and on average, the block itself plus 2 signals\footnote{Distant signal and home signal (German: \textit{Vorsignal} and \textit{Hauptsignal})}, 1 crossing, 1 switch and 2 placeholder elements for the \textit{protection against side collisions}\footnote{Flank protection (German: \textit{Flankenschutz})}, for a total of 7 infrastructure elements. 
Additionally, we have the reservation phase (pre-trip), and the allocation plus deallocation phases during the trip, for a total of 3 phases. 
Thus, we obtain an average of $3*7=21$ write operations per block per train. 

For scenario 1, by applying these estimations, 
we obtain an upper bound indication of
$$2 \times \frac{1}{60} \times 21 = \frac{42}{60} < 1$$ transactions per 
second. 
Thus, it is clear that an enterprise-grade blockchain such as Quorum or Hyperledger Fabric the workload easily: 
a constant stable performance of several thousand transactions per second is reported by independent researchers, e.g. in \cite{baliga2018performance,gorenflo2019fastfabric}. 

For scenario 2, the trains' speeds become more relevant: the higher the speed, the more transactions per second are issued for train control during the train's trip. 
To have a buffer, we assume an unrealistically high average speed of 120km/h, across all trains on that regional network. 
Then, each train covers $\frac{1}{30}$ blocks per second, assuming (as above) that each block has a length of 1000 m. 
With 70 concurrent trains in the network, we obtain a total estimation of $$70\times \frac{1}{30}\times 21 = 49$$ transactions \textit{per second} - still a number that modern DLT can easily cope with. 

For scenario 3, the shorter blocks of 700 m mean that for an average speed of 126 km/h, each train covers $\frac{1}{20}$ blocks per second on average. 
With 300 trains in the network (and using the double number as a buffer), and by assuming the same transaction-per-block approach, we obtain $$600\times \frac{1}{20}\times 21 = 630$$ write transactions \textit{per second} on average. 

Of course, the above workloads do not automatically mean that the performance of the BRCS implementation will be perfectly suitable for all three scenarios. 
A performance analysis needs to account for spikes and load peaks, for queuing and the resulting delays, and also for maximum waiting times in the worst case (e.g. where BRCS has to re-dispatch trains in the case of a disruption). 
Still, we believe that the rough workload estimation shown in this section helps to illustrate that BRCS workloads 
\textcolor{black}{
can be handled by} 
DLT products. 
\textcolor{black}{
Additionally, DLT performance and scalability are continuously improved by the vendors and through the work of the scientific community.} 

\section{Re-Evaluating the Choice of the Blockchain Technology and Execution Platform for BRCS}
\label{ReevaluatingTheChoice}
The modularized architecture of BRCS means that the choice of the specific blockchain technology (and product) does not have a rippling effect on other layers and components. 
Indeed, it is possible to replace the core (safety, allocation and dispatching) while keeping the frontends, and only applying the necessary changes at the interfaces.
Still, the project team decided to perform a systematic analysis before continuing with the implementation of the core. 

\begin{figure}
	[h]
	\begin{center}
		\includegraphics[trim = 45mm 70mm 75mm 15mm, clip, width=0.7\textwidth]
		{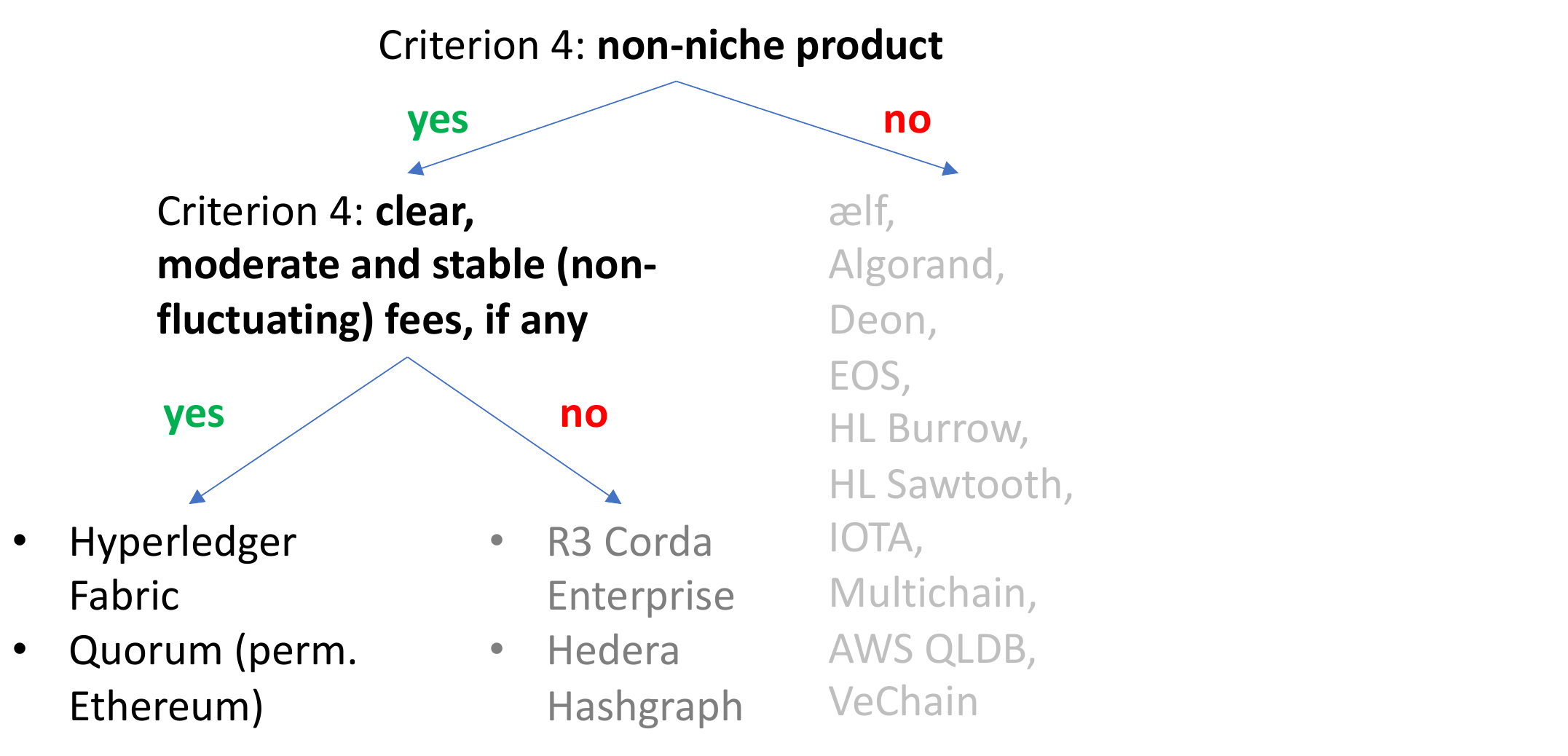}
		\caption{Decision tree for the application of criterion 4 to blockchain product candidates}
		\label{fig-decision-tree}
	\end{center}
\end{figure}

We set up the following list of requirements to guide our decision on whether to keep Ethereum, to switch to Quorum, or to select a different protocol and technology: 
\begin{enumerate}
	\item Performance and scalability requirements of BRCS can be met.

	\item The blockchain network is permissioned and fully IAM-enabled to ensure that only authorized parties can set up nodes, access data and vote. 

	\item Admission to the blockchain network is regulated, e.g. by a subset of participants (e.g. ``authorities'' or ``founders'') or by a majority vote of all/active network participants. 

	\item The blockchain network is based on an established non-niche product; source code inspection and cloud deployment are possible; clear, moderate and stable (non-fluctuating) fees, if any, apply for usage and/or licensing.

	\item Separate/isolated/private test networks can be set up for development, inspection and trials (in-private-cloud and on-developer-machine, before purchase and without commitment). 

	\item The implementation has a broad adaptation base and is backed by a consortium, multi-vendor foundation or interest group (to prevent technology lock-in, product lock-in, vendor lock-in, consultant lock-in etc.). 

	\item Nodeless clients (or light nodes) could run on IT of today's trains, if permitted by regulators. 


	\item Active participation (e.g. voting and block creation) in the network operation is either mandated or incentivized (e.g. through ``gas'' paid for transactions or for mining of blocks); misbehaviour or sabotage can be detected and sanctioned.  

	\item Network participants can find consensus to delete / truncate / rollover data, incl. changes to the source code of smart contracts - deletion would enable GDPR compliance if BRCS needs to store personal data on-chain. 

	\item Support for ``visibility scope limitations'' is present out-of-the-box (similar to Hyperledger Fabric channels). 

	\item API for data access by nodeless clients is available; traffic to and from the nodes is protected at the network protocol level (e.g. using TLS 1.3, which also permits authenticating the client to the server). 

	\item Platform/framework support modularization (at platform/application component level) for simplifying the authorization by oversight agencies and auditors. 

	\item The smart contracts language and runtime are suitable for formal verification or proofs; tools for automated improvements (finding memory leaks, ...) are available. 

	\item Train-initiated communications must support low-bandwith, high-jitter channels. 

	\item Realtime computation guarantees (deterministric execution, QoS for messages and transactions) can be established. 

	\item Channel QoS guarantees for message delivery are provided (e.g. ``at least once'', ``at most once'', etc.).

	\item Visibility control and data isolation facilities are present (e.g. for on-chain payments). 

	\item Transaction management (e.g. using compensations) is possible, preferably with isolation control.

	\item Performance-focused handling of off-chain data is available out-of-the-box. 

\end{enumerate}
We carefully analyzed several DLT products, also taking into consideration the skills and the technology infrastructure of the BRCS project team. 
Fig.~\ref{fig-decision-tree} summarizes the outcome, which we explain below. 

Criterium 4 (``non-niche product'') disqualified several incubating and promising offerings, such as 
aelf, 
Algorand (also due to the vendor lock-in and due to the proprietary cryptocurrency), 
Deon (vendor lock-in, no community), 
EOS (niche), 
Hyperledger Burrow (incubating), 
HL Sawtooth (niche), 
Iota (no smart contracts, partially immature), 
Multichain (no smart contracts), 
AWS QLDB (niche, no decentralization), 
and VeChain (niche).

This left four candidates for the second round: 
Quorum (permissioned Ethereum), 
Hyperledger Fabric (HLF), 
R3 Corda and 
Hedera HashGraph.
R3 Corda Enterprise Edition incurs high costs when deployed in production, whereas Quorum and Fabric have no licensing fees. 
HashGraph charges for mainnet transactions based on its own cryptocurrenty (HBAR), and has a transaction latency of 3-5 seconds; it does not have a free license for private (consortial) network setups. 

This leaves two candidates for the third and final round: Quorum and Fabric. 
As described in \cite{kuperberg2020ledger} and in \cite{kuperberg2019arxiv}, Ethereum was used as the DLT for the original BRCS PoC and showcase. 
When preparing the expansion of BRCS to a real-life prototype, the BRCS team decided to re-visit the suitability of Ethereum, especially with regard to performance. 
Performance analysis requires the quantification of the workload, and how a given setup (DLT runtime and the underlying layers) would perform under this workload. 

In particular, we ran stability and performance tests on Quorum, which is the enterprise-grade version of Ethereum originally developed by JPMorgan. 
Even though Quorum brings design advantages compared to \texttt{geth} (go implementation of Ethereum), Parity, etc., we have found that under high load, reliability of Quorum suffers: for example, we saw legitimate transaction proposals which were not processed. 

In literature, Quorum has been reported to see degraded performance starting at ca. 1900 transactions per second. 
For example, in \cite{baliga2018performance}, Baliga et al. report on transaction latency and throughput of Quorum 2.0, measured for both RAFT and IBFT consensus algorithms using non-virtualized machines with 8 vCPUs and 16 GB RAM each.
In another publication \cite{gorenflo2019fastfabric}, Gorenflo et al. have shown how they have scaled Hyperledger Fabric 1.2 (i.e. an outdated version by 2021) to 20,000 transactions per second, and although they use significantly more powerful servers (24 vCPUs and 64 GB RAM), their setup is also larger: 15 nodes compared to a maximum of 4 nodes in \cite{baliga2018performance}. 

As the result, 
the BRCS team voted to 
move the development to Hyperledger Fabric. 
We believe that it fulfills all requirements (and more requirements than any other candidate); additional requirements engineering and research is necessary to reliably quantify the topics found in the requirements \#14, \#15 and \#16.  


\section{Programming Language and Third-Party Components}
\label{ProgrammingLanguage}
The choice of programming languages for BRCS is constrained by the requirements from rail operations (both EU-wide and national), by the requirements for BRCS, but also by the skills of the development team. 
\textcolor{black}{DIN EN 50128:2011 lists 10 programming languages, classified individually for each of the five safety integrity levels (SIL~0 through SIL~4) into three groups: ``not recommended'', ``recommended'' and ``highly recommended''. 
In Section D.54, this norm also lists some high-level \textit{soft} requirements (worded ``should'', not ``must'') for programming languages.} 
 
As there are no explicit \textcolor{black}{\textit{must}} requirements ``de jure'', BRCS could use the programming languages of the established/pre-existing systems (``de facto'') as a basic assumption and justify its own selection via comparative analyses. 
During the selection progress, we distinguish between ``train-side'' software (including those deliverables that runs on a mobile tablet of a train driver) and the ``land-side'' software. 
In the current BRCS scope, the trains have no blockchain nodes and there are no changes to the train's built-in hardware or software. 

A research report \cite{eba-programmiersprachen} commissioned by the EBA and prepared by Fraunhofer FOKUS, entitled ``Consideration for software development in the railway sector'' analyzes various aspects on more than 60 pages, also in comparison with the automotive industry. 
The research report makes no recommendations for the selection of programming languages; also, it does not object to any programming languages. 
The topic of blockchains/DLT and smart contracts is not dealt with at all. 

BRCS components beyond the blockchain will be written in Java (middleware components, facades etc.) as well in TypeScript (frontend). 
Using conventional project configuration tools (Gradle and Maven), the dependencies are managed across versions, across environments and uniformly for all developers. 

For the ``hard core'' of the business logic within BRCS (the logic is in the smart contracts, called ``chaincode'' in Fabric), after having chosen Hyperledger Fabric, there are 
3 languages to choose from : Java, Go (golang) and TypeScript / JavaScript. 
Unlike R3 Corda, the chaincode is deployed \textit{as-is} - it is not auto-rewritten by the framework. 
The BRCS team chose Java because because automated testing and vulnerability testing (e.g. with JPF) are very mature, but also because the team members are most familiar with this programming language, so better code quality (at a lower cost) can be expected. 
Java is also designed for strong typing (no type inference); the absence of pointer arithmetic is an additional advantage compared to C / C ++. 
``Memory Leaks'' can be found with appropriate tools (such as VisualVM, JProfiler or YourKit) during the test phase (incl. long-term tests and overload tests) and can be fixed accordingly. 
We estimate that the aspects which are rather difficult to implement with Java (e.g. highly parallel, multi-thread logic while avoiding race conditions; transaction management and isolation) will not be present in smart contract logic: these aspects are already covered by Fabric internally and are defined in the framework. 

The other reservations regarding the predictability of applications developed in Java often come from the early days of Java and we consider them as invalid after more than 20 years of development. 
These reservations include the performance and the ``garbage collection'' interruptions as well as the effects of just-in-time compilation. 
The latter is especially relevant as it ensures that the Java bytecode becomes highly performant after a warm-up phase and is almost as fast as native platform-specific code. 
The use of ``Realtime Java'' or of commercial JVMs (instead of OpenJDK) does not currently appear necessary in BRCS. 
During development, the Java runtime components are updated by the team itself (following the DevOps approach, and with regard to published JRE vulnerabilities) and this will later be carried out in cooperation with a platform team that operates the BRCS. 

In addition to the self-developed application components from BRCS, there are also the platform components and third-party libraries (particularly in the frontend layer) that BRCS needs to function. 
In the case of Fabric (which itself is mainly written in Go), the platform components also includes Java execution environments (JRE: Java Runtime Environment; the JVM is part of the JRE as a virtual machine). 
The full inventory of platform components and third-party libraries is compiled by the project management tools such as Maven, and the inventory list can be used to check for vulnerabilities, licensing, duplicates, version conflicts etc.

From the security perspective, one downside of using third-party libraries and platform components is that it would incur disproportionately high resources to manually inspect them for unknown defects and security risks. 
One apparent approach to mitigate these risks is to ``slim down'' these components by removing all functionality that is not used by BRCS. 
However, for many third-party libraries, such a ``shrinking'' requires an extensive and complicated runtime dynamic analysis, which is only conclusive if full code coverage and full coverage of runtime parameters is achieved - a target that is very costly and seldomly achieved. 
Additionally, it would have to be repeated each time a new version of a library appears. 

Licensing is another barrier to altering third-party libraries: in the case of OpenJDK, the GPL license means that altering the rt.jar would require the BRCS project to open-source its changes. 
External shrinking tools such as ProGuard or JShrink\cite{bruce2020jshrink} can optimize the application JARs, but usually leave the library JARs (third-party JARs) unchanged, although researchers have reported that ProGuard can shrink rt.jar \cite{shrink-rt-jar}, too. 

\section{
Safety Integrity Layers (SIL) and Layered Software Architectures}
\label{SafetyIntegrityLayers}
The topics of safety and security and interrelated but distinct: safety (the property of ``being safe'') can be described as the state of being protected from harmful impact, i.e. through control of known threats (``safety measures''). 
Thus, safety is opposed to states of (unmitigated) danger, of (uncontrolled) risk or of (unpredictably) threatful behaviour. 
Software can be made safe even in the presence of \textit{known} defects; \textit{unknown} (``black box'') defects in a component/module can be safeguarded in additional ``safety rings'' around it. 
Statistically, safety is quantified in relation to defined risks (dangers and probabilities). 

Security, on the other side, refers to the processes to secure (literally establish ``freedom from anxiety''), i.e. it refers to procedures and techniques to \textit{protect} by maintaining safety against threats. 
In practice, maintaining safety includes identifying unknown threats, detecting attacks, finding defects and performing similar actions.
Both safety and security are anchored at statistics (probabilities) and are normalized through standards such as laws, norms, best practices) etc. 

It should be noted that there is no ``insurance'' for safety incidents affecting software, unlike for physical assets - in the best case, insurance covers the real-world consequences of such software-affecting incidents. 
At the same time, in the railway domain, safety and security directly affect human lives, which prompts very stringent requirements and processes concerning operations - which has much fewer degrees of freedom than car traffic or bus coach operations. 

A system that is safe must also be reliable and available (the latter requires resiliency, failure tolerance and continuity engineering) - a softwaree system that is not well-maintainable and of low code quality is very unlikely to be safe. 
Also, a system that is not user-friendly is more likely to be less safe - ``idiot-proof'' usability decreases the chances of operating errors; restricting the access to a system makes it more safe (as long as the restriction has measurable effects). 
If a software system remains unaltered, its safety decreases over time (since attackers' arsenals are growing).
Yet at the end, humans are usually the weak spot of any software-using system: humans have a natural ability to violate and to circumvent rules and regulations.

European Norms for Railway Safety and Security are governed by CENELEC (Comité Européen de Normalisation Electrotechnique, assembling 31 states). 
The CENELEC\mbox~norms cover the RAMS ``-ities'': Reliability, Availability, Maintainability and Safety (RAMSST adds Security and Testability, although that acronym is rarely used). 
The CENELEC norms relevant to BRCS include 
EN 50126, 
%
EN 50159, 
EN 50129, 
EN 50128, 
IEC 61508 and others. 

A Safety Integrity Level (SIL) is a classification based on Risk Analysis, SIL 1 being the lowest level and SIL 4 the highest (most dependable).
Quantifying a SIL employs the analysis of systematic errors (systemic errors which are ``baked into'' a system, e.g. due to faulty design or errors in implementation/deployment/sizing) and of random errors (i.e. hardware-caused errors, e.g. due to hardware ageing, hardware faults etc.). 
Systematic errors can trigger random hardware errors, e.g. an overuse of memory hardware through excessive logging \cite{tesla-hardware-errors}. 

Safety of software implies checking the correctness of the software system while striving for the complete coverage of inputs, states and behaviours. 
As software is not ``elastic'', it cannot ``tolerate'' unexpected values and states; additionally, software only regenerates if it is carefully designed to do that. 
If errors are not limited to remain in a submodule, they propagate outwards and to other software modules - effect amplification is often possible
as an error reaches higher software layers without being caught or handled. 
Also, a build-up can happen: errors can lead to outages - and outages can lead to further errors. 
As self-healing software is still to emerge on industrial scale (and so is self-correcting software), \textit{avoidance} of errors (systematic and random) and \textit{reaction} to errors must be balanced based on risks, probabilities and costs. 

SIL levels are based on Tolerable Hazard Rates (THRs), which are specified on a ``per function, per hour'' basis: 
\begin{itemize}
\item SIL 0 denotes ``insecure applications'',
\item SIL 1 is specified as: $10^{-6} < THR < 10^{-5}$
\item SIL 2 is specified as: $10^{-7} < THR < 10^{-6}$
\item SIL 3 is specified as: $10^{-8} < THR < 10^{-7}$
\item SIL 4 is specified as: $10^{-9} < THR < 10^{-8}$
\end{itemize}

To visualize these values, consider an example software system with 100 functions and SIL 4. 
Statistically (!), one has to wait $10^{-7}$ hours or more (i.e. $>416667$ days, i.e. $>1142$ years) until the first (!) systemic software error surfaces. 
Note that the overall error rate might be much higher, given the risk of hardware errors and human errors. 
Also note that this is still an estimation based on heuristics, and also a probabilistic measure!

Thus, while SIL is usable as ``do your best'' guidelines, but it does not provide any guarantees on the outcome. 
In contrast to SIL, Service Level Agreements (SLAs), such as the guaranteed numbers of cloud computing availability (which includes software and hardware!) are defined on the basis of time intervals, in terms of availability (expressed in percentage). 
We therefore believe that the hardware-oriented SIL approach needs a complementing software-centric RAMS approach, especially in the domain of railway safety. 

Research of RAMS and THRs for Blockchain-employing systems needs to consider all employed software layers. 

\section{
The Prospects of Virtualizing the Trackside Equipment for Train Localization and Train Control}
\label{Virtualizing}
In the case of PZB, automated train protection and accompanying in-cab intermittent signaling are implemented through a combination of trackside \textit{oscillating circuit}\footnote{German: \textit{Schwingkreis}} (also called \textit{inductor} or \textit{trackside antenna}) and of trainside equipment (for details on PZB, see e.g. \cite{pachl2018book}). 
PZB (which comes in different variants) is not limited to Germany - it is used in Austria, but also as far as in Israel; in Germany, it is only used for speeds up to 160km/h. 
On modern trains, the trainside physical components of PZB include the counterpart to the trackside oscillating circuit, and an on-board computer that interprets the PZB information and checks the train's speed/braking behaviours and the operator's attention and responsiveness. 

PZB data that the train ``receives'' from the trackside components can be seen as ``pointwise'' events, which come in three types (500 Hz, 1000 Hz, 2000 Hz). 
The computation complexity of how a train reacts to these events is part of the PZB specification, but the events themselves are very simple, as no information about location or other details are submitted - the event can be transmitted as just two bits of information. 
The nature of intermittent signaling means that an absence of events is an ambivalent state: if all parts are functioning correctly then the absence of events it signals a \textit{safe} state but if a component is malfunctioning, absence of events signals an \textit{unsafe} state. 
The physical presense of the oscillating circuit is definitely a reliability factor, but is also a cost factor. 
In fact, trackside PZB can be incorporated into ETCS using so-called STMs (trainside Specific Transmission Modules), which ensure that the trainside EVC (European Vital Computer) can interpret the information that PZB is transmitting. 
Correspondingly, ETCS sees PZB90 as a ``Class~B'' system.  

In BRCS, we have considered the inverse scenario: integrating PZB-equipped vehicles into the BRCS concepts without requiring changes to the pre-existing trainside IT - while also minimizing the trackside PZB IT equipment. 
What is needed to achieve this is a reliable transmission of PZB events into the train's computer without the oscillating circuit, at the right time and in a failsafe way (guaranteed delivery, or detection of connectivity loss). 
A particular challenge is to ensure the timeliness of the signal, especially since conventional hardware-based PZB follows a ``fire and forget'' pattern: if a train fails to read (or to evaluate) the state of the trackside PZB circuit, the trackside circuit will not detect this (no matter if that circuit is functioning properly or not). 

A train moving at 165~km/h covers more than 45 meters per second; event delay (or even event loss) is endangering human lifes. 
If PZB is to be virtualized using wireless communications, it is imperative to ensure a reliable transmission protocol for PZB events, which has to incorporate measures against out-of-order event delivery, multiple delivery etc. 
Unfortunately, even such reliable protocol is not immune against loss of technical connectivity, and we believe a failsafe design for virtualized PZB would involve a train-initiated, periodic ``alive check'' which ensures that the train and the BRCS control infrastructure can communicate in both directions. 
If that communication is interrupted, the train has to be stopped (until further instruction from the dispatcher are received, which is standard provision in PZB) and can only proceed at very low speed, requiring manual overrides from the train driver. 

However, to implement a trainside ``alive check'' means that the original target (making BRCS infrastructure compatible with trainside PZB infrastructure, \textit{without} modifying the trainside IT) requires an \textit{additional} software component that implements the ``safe to proceed only when alive-check successful'' policy. 
On the other hand, modifying the trainside IT infrastructure is a very challenging task, as it requires re-certification of that IT (or developing new IT for the train: software or even hardware). 
Therefore, we have decided not to pursue this idea, and instead to focus on the control and traffic management aspects of BRCS, and we intend to interface BRCS with trackside PZB components (and, implicitly, trainside PZB components). 

Likewise, we considered the compatibility of BRCS with ETCS - at least with Level~1.
Conceptually, ETCS is not as standardized and uniform across countries, as one could assume. 
For example, the ETCS Level~1 standard has a provision for up to 35 (!) values which can be individually configured by each infrastructure undertaking (i.e. network operator); these values are passed to the train as it crosses operator borders (even borders between ETCS operators constitute a complex orchestration \cite{etcs-borders}). 
Still, it is conceptually possible to use BRCS as a control center (``Stellwerk'') which consumes information (e.g. from axle counters, switch controllers, etc.) and provides information, e.g. ETCS Movement Authorizations (MAs) through ETCS RBC (Radio Block Centers), speed restrictions (through ETCS balises or Euroloops) etc. 

However, combining BRCS with trackside and trainside ETCS equipment is hardly a viable business case. 
Additionally, 
many ETCS components are designed to interoperate with preexisting systems using STMs; ETCS Baseline 3 calls the corresponding technology ETCS Level~NTC (National Train Control) and positions ETCS Level~NTC between Level~0 (e.g. no train control; trainside ETCS controls max. speed only) and ETCS Level~1.

\section{Quick Release Cycles}
\label{QuickReleaseCycles}
\textcolor{black}{
Whether for bugfixes or for feature upgrades, the BRCS approach will need to accomodate the capability for installing a later release 
as a replacement of a given, working installation. 
Enterprise-grade software is subject to thorough planning and documentation of releases, including feature set, migration policy, backporting of bugfixes, etc. 
For BRCS, drafting 
the policy of how releases are defined, managed and deployed is an important question because in a distributed system with multiple active nodes, maintaining a good state of the network (and data integrity) depends on how releases and updates are deployed onto the nodes.}

\begin{figure}
	[h]
	\begin{center}
		\includegraphics[trim = 11mm 76mm 120mm 21mm, clip, width=0.7\textwidth]
		{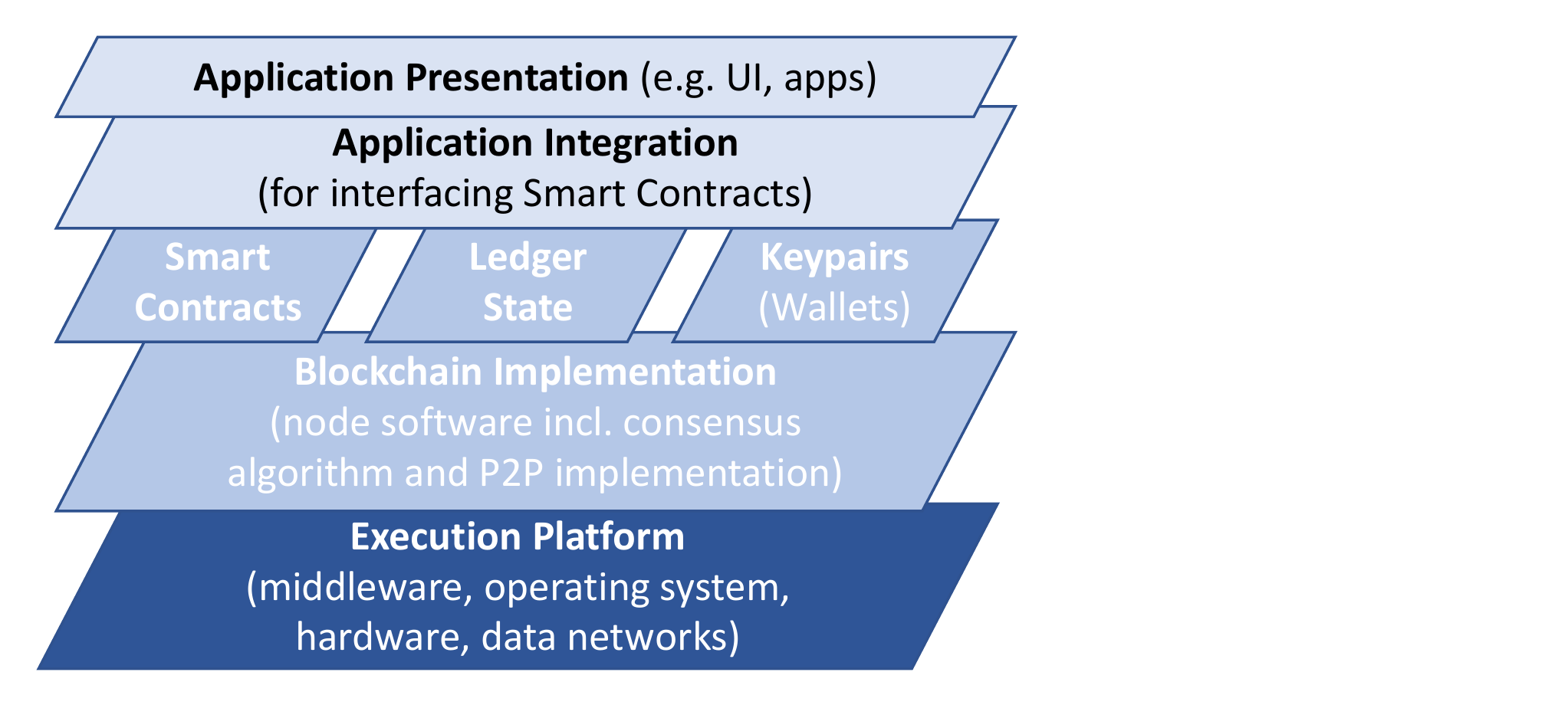}
		\caption{Layers of a blockchain application incl. the runtime platform}
		\label{fig-layers}
	\end{center}
\end{figure}

A release may concern the application components or the platform components, and a release impacts not just code (behaviour) but also the stored data (state of the application and of the platform).
As shown in Fig.~\ref{fig-layers}, BRCS will contain both externally-sourced layers and project-created layers.  
Inconsistencies in complex, layered and componentized applications can be prevented using systematic configuration management and acceptance testing (even for ``standard'', externally-sourced COTS dependencies). 
To save time and space, updates are often designed to be incremental or differential - 
at the expense of having to ensure that such updates are applied in correct order, atomically and in a durable way. 

\subsection{Specifics of Railway Control Software}
\label{QuickReleaseCycles-Specifics}
In railway control, software updates and upgrades impact operations (often requiring complete service suspension) and are therefore frequently bundled with hardware upgrades or regular infrastructure maintenance. 
Thus, a release may be a critical, disruptional bugfix or it can be a non-essential release which can be scheduled based on priorities, traffic, workforce availability, etc. 
Unlike in consumer-oriented applications, safety-relevant updates require utmost care, and a ``hotfix'' mentality is not accepted. 
Thus, a critical bugfix must be issued as a qualified quick release: not on a regular schedule, but still with sufficient testing. 	
However, systematic software development is structured into processes and the higher the share of non-automated testing, the more time will pass between scope freezing and actual release. 
Even in DevOps (and in DevSecOps), user acceptance testing involves planning and is constrained by availability of skilled personnel. 

A 
straightforward
approach to interruption-minimizing release \textit{deployment} is the \textit{side-by-side operation} of the established release alongside the new release. 
Such setup is initiated by seeding the new release with a valid snapshot of data from the established release, and then sending all requests to \textit{both} installations.
By comparing the new release's behaviour and data to the expected one, the deployment can be validated and eventually, the new release can be promoted to ``definitive'' and the old release can be downgraded to ``deprecated''. 
However, such an approach is both complex and costly: often, the implementation and the behaviour change to such a degree that the two releases diverge, and converters are needed for seeding and to transport the data to both releases, or between them.

Additionally, such setup requires that the pre-existing setup is shielded from new release in terms of side effects and visibility: for example, in messaging-based middleware, message brokers duplicate a message $M_1$ from sender $A$ to release $X$ and send it to the queue of release $Y$ - without $A$ being aware of the duplication. 
While this sounds intuitive and doable, the practical realization can be a nightmare if interface protocols change. 
For example, if $M_1$ triggers a request-response exchange from $X$ to $A$ (i.e. $M_{req}$, $M_{resp}$), than $X$ awaits $M_{resp}$ and will have some error handling if $M_{resp}$ does not arrive or is delayed. 
However, if $Y$ is run alongside $X$ but expects a different response $M_{resp'}$, duplicating $M_{resp}$ in the message broker and sending $M_{resp}$ to $Y$ will cause $Y$ to detect an error. 
At the same time, it would be a waste of efforts to create a custom message handler in the broker to adapt the cloned message $M_{resp}$ to the format $M_{resp'}$ expected by $Y$. 

Generally, running two releases at the same time requires coordination - even more so in multi-party deployment such as blockchain business networks, or when a distributed deployment must support compatible versions. 
If enforced stringently, a multi-node setup may be streamlined to disallow 
different releases, even if they would be compatible or interoperable. 
Choosing between availability and consistency, it may be wise to prefer consistency over availability when safety is an imperative. 

Specifically for railway control (i.e. unlike in software-only settings), even when test-running different releases in parallel, only one release can in general be enabled to interact with trackside hardware and trainside hardware. 
Thus, acceptance testing of a new release can only happen when normal operations are interrupted. 
Therefore, when following the quick release cycles approach, a railway control system may not follow a laissez-faire approach to test coverage: it must incorporate state-of-the-art testing and QA methodologies for keeping quick 
releases covered efficiently and sufficiently. 

\subsection{DLT-focused Aspects}
\label{QuickReleaseCycles-DLT}
At first glance, blockchain-using systems can follow the known strategies when attempting to implement quick release cycles. 
Simply spoken, encapsulation/modularization and interface-level contracts are essential, and control of side effects (e.g. performance isolation between different modules) is definitely a critical success factor. 
\textcolor{black}{Unlike in a traditional ``vendor-buyer'' relationship, releases of blockchain-deployed software must be coordinated within the blockchain business network, i.e. between several (and possibly not only cooperating, but competing) organizations.} 

As mentioned above, test automation is another contributing factor and formal methods, if appropriate, can add further assurance. 
Making undertested functionality temporarily unavailable (until a later release where sufficient test coverage would make it available again) may work for some edge cases, but requires a dependency analysis of its own and is unlikely to be popular with customers and authorizing agencies. 

However, \textit{if} rollbacks of releases have to be supported (as a mitigation strategy should a quick release be pushed out \textit{too} quickly), additional intricacies need to be considered. 
Additionally, there is always the (undesired) option of avoiding a rollback from Y to X and instead either supplementing the faulty quick release Y with another quick release Z. 
Note that \textit{if} Z is incremental (based on Y) rather than differential (based on X), taking the last valid snapshot based on X and \textit{directly} applying Z is not possible.
Instead, an update Z' based on X has to be created to bypass Y - which results in even more implementation and testing.  

Unlike ``redo logs'' known from RDBMS, consensus-based ledgers do not have an ``administrator'' which could re-fill the system with history of transactions (and who could modify that history if necessary). 
Depending on a specific ledger implementation, transaction proposals (but also endorsements, and sealed blocks) may be timestamped using a trusted third-party timestamping service which uses cryptography, and later blocks may reference those timestamps. 
While snapshotting functionality \cite{fabric-snapshots} has been included in recent releases of Hyperledger Fabric, it is not meant for infrastructure upgrades or rollbacks, but rather for quick bootstrapping of nodes and for slimming the storage requirements. 
Also, recovering a Fabric state from a backup requires coordination between data backup, configuration backup (incl. CAs and PKI artefacts) and verification with organizational updates. 

Based on the considerations outline above, the following aspects are 
initial thoughts
which need to be validated in practical settings. 
\begin{itemize}
	\item \textbf{Append-only semantics in DLT}: rolling back a release Y to release X \textit{may} be subject to data incompatibility, if the newer release Y uses a different data \textit{format} from that in X. As data in hash-concatenated append-only ledgers is signed, \textit{if} storage-level data format differences affect the signed contents of a transaction/block, the situation arises where we cannot simply ``copy'' data from Y to X: the append-only ledger would have to be rebuilt to match the formatting in X, including block validations. In Hyperledger Fabric, blocks are validated following an endorsement policy, and may involve more than one organization from the given Fabric network. 
	\item \textbf{Validity of certificates}: if previously endorsed transactions were signed with then-valid certificates which have expired by the time where transaction history has to be replayed (during an upgrade or during a rollback), the DLT implemetnation's behaviour has to be studied. 
	\item \textbf{Changes in network membership and endorsement policies}: when an organization leaves the network, replaying a transaction history is subject to the same questions as when a certificate / private key has expired. 
	\item \textbf{Transaction queuing}: during the system downtime (for an upgrade or for a rollback), transaction requests may continue to come in if the transaction-submitting system is not aware of the downtime, or if decoupling is a design choice to minimize effects of downtimes. However, queued DLT-bound transactions may become outdated by the time when the system is running again. It remains to be studied how Hyperledger Fabric reacts to transactions which have been correct at the time of submission, but become outdated after an outage (for example, booking a route for a timeslot which has since become unavailable). 
	\item \textbf{Non-determinism and time-dependent behaviour in chaincode}: smart contracts are not required to be fully deterministic or idempotent. Thus, if transactions need to be replayed, the assumptions need to be validated and documented. 
	\item Unlike mass market software, BRCS adopter are unlikely to group into ``early adopters'' and ``only after it has settled'' conservatives. Instead, customers will expect a release to be hardened and fully quality-assured. 
\end{itemize}

Beyond these specifics, well-known difficulties of the quick release cycles approach remain, among them
\begin{itemize}
	\item \textbf{Less time for saturation testing}: certain system behaviour may only emerge after running a system for sufficient time. For example, relational databases are known for malfunctioning once the ``archive log'' (redo log) is full, even if the main storage space has space left. If we consider configuration as part of implementation, the specific configuration settings must be tested alongside application code, and corrected if needed. 
	\item \textbf{``Stopgap code''} that is a temporary fill-in and which is slated to be removed by ``proper'' implementation in the next release. Thus, the stopgap code is often under-tested or over-rushed because the resources are allocated to the ``proper'' implementation of that functionality that will be released in a later release. 
	\item \textbf{Dependency management}, i.e. keeping componentized applications well-structured and free of cyclic dependencies requires constant monitoring of how the architecture is implemented. Striking a balance between small-but-many (``microservices'' approach, with associated efforts for orchestration and managing availability despite loose coupling) and few-but-large (``oligoliths'': large components which are cleanly modularized internally, but can only be deployed atomically as individual monoliths) can be very challenging. 
	\item \textbf{Tests may be too simple despite full source code coverage}: Many developers prefer to focus on unit tests, because these require less coordination and are conceptually simpler (even at the expense of having to create mocks for isolation testing). In contrast to that, integration tests across components, developers or even teams are intrinsically complex and ad-hoc changes will break the tests (unless a test-first approach is followed). 
\end{itemize}

\section{Conclusions}
\label{Conclusions}
In this \textcolor{black}{technical report}, we have discussed the progress of BRCS that has been made since \cite{kuperberg2020ledger}. 
In particular, we have explained how the the BRCS team has re-evaluated its choice of the DLT product, and why we switched to Hyperledger Fabric. 
New insights affecting reliability and scalability include an analysis of network partitioning, and a new architecture for the ledger, where the latter paves the road to a deletion-enabled append-only DLT, and helps contain data growth of on-chain data. 

Business-wise, we have described the potential customers and markets for BRCS, and provided an insight into the workloads and how they translate into the transaction arrival behaviour for the DLT. 
The model-based analysis of DLT workloads shows that BRCS workloads for both regional networks and heavily used corridors are feasible scenarios for our architecture. 
Additional technical details covered in this paper include the choice of the programming language for smart contracts, the applicability of Safety Integrity Levels (SILs) as measures of RAMS, and the compatibility with prexisting trainside and trackside infrastructure. 

As of February 2021, BRCS is establishing the next round of funding to continue its research work. 
This coincides with the largest challenge that BRCS is facing: coordinating the acceptance of safety-relevant railway control software with government authorities (EBA) and infrastructure undertakings (such as DB Netz) requires an intensive long-term collaboration as well as CENELEC-mandated development and documentation processes, and suggested budgets for this setup are seven-figure amounts. 
Such a budget is only viable if enough purchasing customers sign up even before BRCS development is finished, and is deterred by the law-mandated choice of ETCS as long-term train control technology in the EU. 
Additionally, ETCS has a decade-long forerun and a track record of deployed installations - yet we believe that BRCS does have a viable market given the excessive costs and suboptimal ROI for low-traffic lines. 

Next research steps in BRCS include the quantification of availability of the BRCS software stack, and the study of scalability based on implemented Hyperledger Fabric chaincode. 
Additionally, we plan to build a simulator to study end users' experience with the BRCS workflows. 

\section*{Acknowledgements}

This \textcolor{black}{technical report} describes work performed by a team at DB Systel GmbH, and not just the author 
of this paper. 
After the prototyping/showcase phase, the following individuals (in alphabetical order) contributed code and other artefacts to the work on BRCS, in different roles and over different timespans:
Philipp Bletzer, 
Ngoc-Giao Han,
Annabella Kakur,
Daniel Kindler,
Koraltan Kaynak,
Sebastian Kemper, 
Roland Kittel,
Daniel Kolb,
Andre Mendzigall, 
Christian Müller,
Frank Murrmann, 
\textcolor{black}{Mahoma Niemeyer}, 
Malte Plescher,
Anna Schoderer,
Josefine Schumann,
Jan Schwaibold,
Maximilian Skarzynski, 
Marta Smuk
and 
\textcolor{black}{Dawid Tunkel}. 
\textcolor{black}{Norbert Just and Robin Klemens provided helpful review comments.} 
Further individuals (mostly from the Blockchain\&DLT group at DB Systel GmbH) engaged in helpful discussions. 
The BRCS patent authors laid the cornerstone. 

\bibliographystyle{IEEEtran}
\bibliography{Kuperberg2021a-BRCS-Arxiv}

\begin{thebibliography}{10}
\providecommand{\url}[1]{#1}
\csname url@samestyle\endcsname
\providecommand{\newblock}{\relax}
\providecommand{\bibinfo}[2]{#2}
\providecommand{\BIBentrySTDinterwordspacing}{\spaceskip=0pt\relax}
\providecommand{\BIBentryALTinterwordstretchfactor}{4}
\providecommand{\BIBentryALTinterwordspacing}{\spaceskip=\fontdimen2\font plus
\BIBentryALTinterwordstretchfactor\fontdimen3\font minus
  \fontdimen4\font\relax}
\providecommand{\BIBforeignlanguage}[2]{{%
\expandafter\ifx\csname l@#1\endcsname\relax
\typeout{** WARNING: IEEEtran.bst: No hyphenation pattern has been}%
\typeout{** loaded for the language `#1'. Using the pattern for}%
\typeout{** the default language instead.}%
\else
\language=\csname l@#1\endcsname
\fi
#2}}
\providecommand{\BIBdecl}{\relax}
\BIBdecl

\bibitem{etcs-official}
\BIBentryALTinterwordspacing
{European Union Agency for Railways (ERA)}. {CCS TSI Annex A – Mandatory
  specifications}. [Online]. Available:
  \url{https://www.era.europa.eu/content/ccs-tsi-annex-mandatory-specifications}
\BIBentrySTDinterwordspacing

\bibitem{ertms-official}
\BIBentryALTinterwordspacing
------. {European Rail Traffic Management System (ERTMS)}. [Online]. Available:
  \url{https://www.era.europa.eu/activities/european-rail-traffic-management-system-ertms\_en}
\BIBentrySTDinterwordspacing

\bibitem{KuperbergKindlerJeschke2020}
\BIBentryALTinterwordspacing
M.~Kuperberg, D.~Kindler, and S.~Jeschke, ``{Are Smart Contracts and
  Blockchains Suitable for Decentralized Railway Control?}'' \emph{Ledger},
  vol.~5, Aug. 2020. [Online]. Available:
  \url{http://www.ledgerjournal.org/ojs/ledger/article/view/158}
\BIBentrySTDinterwordspacing

\bibitem{holbrook2020book}
J.~Holbrook, \emph{{Architecting Enterprise Blockchain Solutions}}.\hskip 1em
  plus 0.5em minus 0.4em\relax Wiley, 2020.

\bibitem{lorne2020book}
L.~Lantz and D.~Cawrey, \emph{{Mastering Blockchain: Unlocking the Power of
  Cryptocurrencies, Smart Contracts, and Decentralized Applications}}.\hskip
  1em plus 0.5em minus 0.4em\relax {O'Reilly}, 2020.

\bibitem{maschek2018buch}
U.~Maschek, \emph{{Sicherung des Schienenverkehrs}}.\hskip 1em plus 0.5em minus
  0.4em\relax {Springer Vieweg}, 2018.

\bibitem{pachl2018book}
J.~Pachl, \emph{{Railway Operation and Control}}.\hskip 1em plus 0.5em minus
  0.4em\relax {VTD Rail Publishing}, 2018.

\bibitem{kuperberg2020ledger}
\BIBentryALTinterwordspacing
M.~Kuperberg, D.~Kindler, and S.~Jeschke, ``{Are Smart Contracts and
  Blockchains Suitable for Decentralized Railway Control?}'' \emph{LEDGER},
  (Journal Article) 2020. [Online]. Available:
  \url{http://www.ledgerjournal.org/ojs/ledger/article/view/158}
\BIBentrySTDinterwordspacing

\bibitem{androulaki2018hyperledger}
\BIBentryALTinterwordspacing
E.~Androulaki, A.~Barger, V.~Bortnikov, C.~Cachin, K.~Christidis, A.~De~Caro,
  D.~Enyeart, C.~Ferris, G.~Laventman, Y.~Manevich, S.~Muralidharan, C.~Murthy,
  B.~Nguyen, M.~Sethi, G.~Singh, K.~Smith, A.~Sorniotti, C.~Stathakopoulou,
  M.~Vukoli\'{c}, S.~W. Cocco, and J.~Yellick, ``{Hyperledger Fabric: A
  Distributed Operating System for Permissioned Blockchains},'' in
  \emph{Proceedings of the Thirteenth EuroSys Conference}, ser. EuroSys
  '18.\hskip 1em plus 0.5em minus 0.4em\relax New York, NY, USA: Association
  for Computing Machinery, 2018. [Online]. Available:
  \url{https://doi.org/10.1145/3190508.3190538}
\BIBentrySTDinterwordspacing

\bibitem{tradelens}
\BIBentryALTinterwordspacing
{GTD Solution Inc. and IBM (together the “TradeLens Collaboration”)}.
  Tradelens. [Online]. Available: \url{https://www.tradelens.com}
\BIBentrySTDinterwordspacing

\bibitem{jensen2019tradelens}
T.~Jensen, J.~Hedman, and S.~Henningsson, ``{How TradeLens Delivers Business
  Value With Blockchain Technology.}'' \emph{MIS Quarterly Executive}, vol.~18,
  no.~4, 2019.

\bibitem{kuperberg2018stationshalte}
\BIBentryALTinterwordspacing
M.~Kuperberg, P.~Sandner, and M.~Felder, ``{Blockchain-basierte Abrechnung der
  IoT- registrierten Stationshalte: ein Proof-of-Concept auf Basis von
  Ethereum},'' Frankfurt School of Business and Finance, Blockchain Center,
  Tech. Rep., 2018. [Online]. Available: \url{https://bit.ly/3isB2Y9}
\BIBentrySTDinterwordspacing

\bibitem{rca-eulynx}
\BIBentryALTinterwordspacing
{EULYNX}. {RCA Baseline set 0 Release 1 Update (04.02.2021)}. [Online].
  Available: \url{https://public.3.basecamp.com/p/KeehzqFmXv5R2N7tGDjaEokq}
\BIBentrySTDinterwordspacing

\bibitem{ocora}
\BIBentryALTinterwordspacing
{EULYNX Consortium}. Ocora. [Online]. Available:
  \url{https://github.com/OCORA-Public/}
\BIBentrySTDinterwordspacing

\bibitem{dsd}
\BIBentryALTinterwordspacing
DSD. {Digitale Schiene Deutschland}. [Online]. Available:
  \url{https://www.digitale-schiene-deutschland.de/en}
\BIBentrySTDinterwordspacing

\bibitem{shift2rail}
\BIBentryALTinterwordspacing
{EIM}. {Shift2Rail Successor}. [Online]. Available:
  \url{https://shift2rail.org/shift2rail-successor/}
\BIBentrySTDinterwordspacing

\bibitem{db-wettbewerbsbericht-fuer-2018-2019}
\BIBentryALTinterwordspacing
{DB AG}, \emph{{Wettbewerbskennzahlen 2018/2019}}.\hskip 1em plus 0.5em minus
  0.4em\relax {DB AG}, 2020. [Online]. Available:
  \url{https://www.deutschebahn.com/resource/blob/4593160/52024c17f17fd
  809cd4c9c9a58de1954/Wettbewerbskennzahlen-2018\_19-data.pdf}
\BIBentrySTDinterwordspacing

\bibitem{liefferinge2017-etcs-compatibility}
\BIBentryALTinterwordspacing
M.~V. Liefferinge. {PROPOSALS ON ETCS COMPATIBILITY TESTING AND
  RE-AUTHORIZATION}. [Online]. Available:
  \url{https://www.era.europa.eu/sites/default/files/events-news/docs/ccrcc\_2017\_proposals\_for\_testing\_and\_re-authorizationl\_unisig\_en.pdf}
\BIBentrySTDinterwordspacing

\bibitem{karg2016openetcs}
\BIBentryALTinterwordspacing
S.~Karg, A.~Raschke, M.~Tichy, and G.~Liebel, ``Model-driven software
  engineering in the openetcs project: Project experiences and lessons
  learned,'' in \emph{Proceedings of the ACM/IEEE 19th International Conference
  on Model Driven Engineering Languages and Systems}, ser. MODELS '16.\hskip
  1em plus 0.5em minus 0.4em\relax New York, NY, USA: Association for Computing
  Machinery, 2016, p. 238–248. [Online]. Available:
  \url{https://doi.org/10.1145/2976767.2976811}
\BIBentrySTDinterwordspacing

\bibitem{berendea2020fair}
N.~Berendea, H.~Mercier, E.~Onica, and E.~Rivière, ``{Fair and Efficient
  Gossip in Hyperledger Fabric},'' 2020.

\bibitem{6133253}
E.~{Brewer}, ``{CAP twelve years later: How the "rules" have changed},''
  \emph{Computer}, vol.~45, no.~2, pp. 23--29, Feb 2012.

\bibitem{alquraan2018-partitioning}
\BIBentryALTinterwordspacing
A.~Alquraan, H.~Takruri, M.~Alfatafta, and S.~Al-Kiswany, ``{An Analysis of
  Network-Partitioning Failures in Cloud Systems},'' in \emph{13th {USENIX}
  Symposium on Operating Systems Design and Implementation ({OSDI} 18)}.\hskip
  1em plus 0.5em minus 0.4em\relax Carlsbad, CA: {USENIX} Association, Oct.
  2018, pp. 51--68. [Online]. Available:
  \url{https://www.usenix.org/conference/osdi18/presentation/alquraan}
\BIBentrySTDinterwordspacing

\bibitem{kuperberg2020partitioning}
\BIBentryALTinterwordspacing
M.~Kuperberg, ``{Towards An Analysis of Network Partitioning Prevention for
  Distributed Ledgers and Blockchains},'' in \emph{The 2nd IEEE International
  Conference on Decentralized Applications and Infrastructures}, P.~Ruppel,
  S.~Schulte, and D.~Jadav, Eds., 2020. [Online]. Available:
  \url{https://ieeexplore.ieee.org/document/9126030}
\BIBentrySTDinterwordspacing

\bibitem{clique}
\BIBentryALTinterwordspacing
P.~Szilágyi. {EIP-225: Clique proof-of-authority consensus protocol, Ethereum
  Improvement Proposals, no. 225}. [Online]. Available:
  \url{https://eips.ethereum.org/EIPS/eip-225}
\BIBentrySTDinterwordspacing

\bibitem{ibft10}
\BIBentryALTinterwordspacing
Y.-T. Lin. {Istanbul Byzantine Fault Tolerance}. [Online]. Available:
  \url{https://github.com/ethereum/EIPs/issues/650}
\BIBentrySTDinterwordspacing

\bibitem{2019arXiv190910194S}
R.~{Saltini} and D.~{Hyland-Wood}, ``{{IBFT 2.0: A Safe and Live Variation of
  the IBFT Blockchain Consensus Protocol for Eventually Synchronous
  Networks}},'' \emph{arXiv e-prints}, p. arXiv:1909.10194, Sep 2019.

\bibitem{kuperberg2020deletion}
\BIBentryALTinterwordspacing
M.~Kuperberg, ``{Towards Enabling Deletion in Append-Only Blockchains to
  Support Data Growth Management and GDPR Compliance},'' in \emph{2020 IEEE
  International Conference on Blockchain (Blockchain)}, Nov 2020, pp. 393--400.
  [Online]. Available: \url{https://ieeexplore.ieee.org/document/9284810}
\BIBentrySTDinterwordspacing

\bibitem{kuperberg2020arxiv}
\BIBentryALTinterwordspacing
------, ``{Enabling Deletion in Append-Only Blockchains (Short Summary / Work
  in Progress)},'' DB Systel GmbH / arXiv, Tech. Rep., 2020. [Online].
  Available: \url{https://arxiv.org/abs/2005.06026}
\BIBentrySTDinterwordspacing

\bibitem{DE1020182202249}
\BIBentryALTinterwordspacing
------, ``{Verfahren zum manipulationssicheren Speichern von Daten in einem
  elektronischen Speicher unter Verwendung einer verketteten
  Blockchain-Struktur}.'' [Online]. Available:
  \url{https://register.dpma.de/DPMAregister/pat/register?AKZ=
  1020182202249\&CURSOR=0}
\BIBentrySTDinterwordspacing

\bibitem{baliga2018performance}
\BIBentryALTinterwordspacing
A.~Baliga, I.~Subhod, P.~Kamat, and S.~Chatterjee, ``{Performance Evaluation of
  the Quorum Blockchain Platform},'' 2018. [Online]. Available:
  \url{https://arxiv.org/abs/1809.03421}
\BIBentrySTDinterwordspacing

\bibitem{gorenflo2019fastfabric}
C.~Gorenflo, S.~Lee, L.~Golab, and S.~Keshav, ``{FastFabric: Scaling
  Hyperledger Fabric to 20,000 Transactions per Second},'' 2019.

\bibitem{kuperberg2019arxiv}
\BIBentryALTinterwordspacing
M.~Kuperberg, D.~Kindler, and S.~Jeschke, ``{Are Smart Contracts and
  Blockchains Suitable for Decentralized Railway Control?}'' DB Systel GmbH /
  arXiv, Tech. Rep., 2019. [Online]. Available:
  \url{http://arxiv.org/abs/1901.06236}
\BIBentrySTDinterwordspacing

\bibitem{eba-programmiersprachen}
\BIBentryALTinterwordspacing
{Fraunhofer FOKUS / Eisenbahnbundesamt}. Consideration of software development
  in the railway sector. [Online]. Available:
  \url{https://www.bmvi-expertennetzwerk.de/DE/Themen/Themenfeld4/themenfeld4\_node.html}
\BIBentrySTDinterwordspacing

\bibitem{bruce2020jshrink}
\BIBentryALTinterwordspacing
B.~R. Bruce, T.~Zhang, J.~Arora, G.~H. Xu, and M.~Kim, \emph{{JShrink: In-Depth
  Investigation into Debloating Modern Java Applications}}.\hskip 1em plus
  0.5em minus 0.4em\relax New York, NY, USA: Association for Computing
  Machinery, 2020, p. 135–146. [Online]. Available:
  \url{https://doi.org/10.1145/3368089.3409738}
\BIBentrySTDinterwordspacing

\bibitem{shrink-rt-jar}
\BIBentryALTinterwordspacing
J.~Palmer. {rt.jar put on a Proguard diet losses 26 MB}. [Online]. Available:
  \url{http://mail.openjdk.java.net/pipermail/macosx-port-dev/2012-December/005214.html}
\BIBentrySTDinterwordspacing

\bibitem{tesla-hardware-errors}
\BIBentryALTinterwordspacing
N.~Broekhuijsen. {Flash Memory Wear Killing Older Tesla's Due to Excessive Data
  Logging: Report}. [Online]. Available:
  \url{https://www.tomshardware.com/news/flash-memory-wear-killing-older-teslas-due-to-excessive-data-logging-report}
\BIBentrySTDinterwordspacing

\bibitem{etcs-borders}
\BIBentryALTinterwordspacing
D.~B. AG. {Transitionen und Grenzübergänge}. [Online]. Available:
  \url{https://fahrweg.dbnetze.com/fahrweg-de/kunden/nutzungsbedingungen/etcs/fahrzeuganforderungen/etcs\_ntr-2080742}
\BIBentrySTDinterwordspacing

\bibitem{fabric-snapshots}
\BIBentryALTinterwordspacing
H.~Fabric. {Taking ledger snapshots and using them to join channels}. [Online].
  Available:
  \url{https://hyperledger-fabric.readthedocs.io/en/latest/peer\_ledger\_snapshot.html}
\BIBentrySTDinterwordspacing

\end{thebibliography}

\end{document}